\RequirePackage{etoolbox}
\csdef{input@path}{%
 {sty/}
 {img/}
}%

\documentclass[ba]{imsart}
\pubyear{0000}
\volume{00}
\issue{0}
\doi{0000}
\firstpage{1}
\lastpage{1}

\usepackage{amsthm}
\usepackage{amsmath,amssymb,bm,bbm}
\usepackage{natbib}
\usepackage{apalike}
\usepackage{mathtools}

\usepackage[modulo,pagewise]{lineno}

\usepackage{float}
\usepackage{graphicx,psfrag,epsf}
\usepackage{enumerate}
\usepackage{dsfont}
\usepackage{subcaption}
\usepackage{url} 
\usepackage{geometry}
\usepackage{marginnote}
\usepackage{xcolor}
\usepackage[utf8]{inputenc}
\usepackage{comment}
\usepackage[colorinlistoftodos]{todonotes}

\newcommand{\N}{\mathrm{N}}
\newcommand\iidsim{\stackrel{\mathrm{  i.i.d.}}{\sim}}

\startlocaldefs
\numberwithin{equation}{section}
\theoremstyle{plain}
\newtheorem{thm}{Theorem}
\newtheorem{definition}{Definition}
\newtheorem{corollary}{Corollary}
\endlocaldefs

\begin{document}

\begin{frontmatter}
\title{Bayesian Inference in Nonparanormal Graphical Models}
\runtitle{Bayesian Nonparanormal Graphical Models}

\begin{aug}
\author{\fnms{Jami J.} \snm{Mulgrave}\thanksref{addr1,t1}\ead[label=e1]{jnjacks3@ncsu.edu}}
\and
\author{\fnms{Subhashis} \snm{Ghosal}\thanksref{addr2}\ead[label=e2]{sghosal@ncsu.edu}}


\runauthor{J. J. Mulgrave \& S. Ghosal}

\address[addr1]{North Carolina State University, Department of Statistics
    \printead{e1} 
}

\address[addr2]{North Carolina State University, Department of Statistics
    \printead*{e2}
}


\end{aug}

\begin{abstract}
Gaussian graphical models have been used to study intrinsic dependence among several variables, but the Gaussianity assumption may be restrictive in many applications. A nonparanormal graphical model is a semiparametric generalization  for continuous variables where it is assumed that the variables follow a Gaussian graphical model only after some unknown smooth monotone transformations on each of them. We consider a Bayesian approach in the nonparanormal graphical model by putting priors on the unknown transformations through a random series based on B-splines where the coefficients are ordered to induce monotonicity.  A truncated normal prior leads to partial conjugacy in the model and is useful for posterior simulation using Gibbs sampling. On the underlying precision matrix of the transformed variables, we consider a spike-and-slab prior and use an efficient posterior Gibbs sampling scheme. We use the Bayesian Information Criterion to choose the hyperparameters for the spike-and-slab prior. We present a posterior consistency result on the underlying transformation and the precision matrix. We study the numerical performance of the proposed method through an extensive simulation study and finally apply the proposed method on a real data set.
\end{abstract}



\end{frontmatter}

\section{Introduction}

Graphical models describe intrinsic relationships among a collection of variables. Each variable in the collection is represented by a node or a vertex. Two nodes in the graph are connected by an edge if and only if the corresponding variables are not conditionally independent given the remaining variables.  Conditional independence impacts the precision matrix, that is, the inverse  covariance matrix, by setting the $(i,j)$th off-diagonal entry to zero if the random variables associated with the $i$th and $j$th nodes are conditionally independent given others. Conditional independence makes the partial correlation coefficient between the random variables associated with the $i$th and $j$th entries equal to zero as well.  If the random variables in the collection can be assumed to be jointly normally distributed, then the conditional independence between the $i$th and the $j$th variables is exactly equivalent to having the $(i,j)$th entry of the precision matrix equal to zero. Such models are known as Gaussian Graphical Models (GGMs).  Learning the conditional dependence structure in a GGM is therefore equivalent to estimating the corresponding precision matrix under the assumed sparsity condition. Modeling intrinsic dependence between random variables through GGMs is commonly used in biology, finance, and the social sciences.

Estimation of a sparse precision matrix needs some form of regularization.  In the non-Bayesian literature, the estimation is typically carried out by minimizing the penalized log-likelihood of the data with the $\ell_1$-penalty on the elements of the precision matrix.  This method is known as the graphical lasso  \citep{friedman_sparse_2008}. Many algorithms have been proposed to solve this problem  \citep{meinshausen_high-dimensional_2006,yuan_model_2007,friedman_sparse_2008, banerjee_model_2008, daspremont_first-order_2008,rothman_sparse_2008, lu_smooth_2009, scheinberg_sparse_2010, witten_new_2011,mazumder_graphical_2012}. 

Bayesian methods for GGMs involve using priors on the precision matrix and priors on the graph as well. A popular prior on a precision matrix is given by the family of G-Wishart priors \citep{giudici_decomposable_1999, letac_wishart_2007,wang_efficient_2012}. The G-Wishart prior is conjugate to multivariate normal random variables and yields an explicit expression for the posterior mean. If the underlying graph is decomposable, the normalizing constant in a G-Wishart distribution has a simple closed form expression. In the absence of decomposability, the expression is more complex \citep{uhler_exact_2018}, but may be computed by simulations.  Simulations from a G-Wishart distribution is possible using the R package {\tt BDgraph} \citep{mohammadi_bdgraph:_2017,
mohammadi_bdgraph:_2019}, which uses an explicit expression for the normalizing
constant for a decomposable graph and uses the birth-death MCMC algorithm \citep{mohammadi_bayesian_2015} if the graph is not decomposable. This allows computation of the marginal likelihood, and hence the posterior probability, of any given graph. However, as the number of possible graphs is huge, computing posterior probabilities of all graphs is an impossible task for even a modest number of nodes. The problem is worsened by the fact that a very low fraction of graphs are decomposable. Thus when learning the graphical structure from the data, alternative mechanisms of putting priors on the entries of the precision matrix that allow sparsity are typically employed. A prior that models a sparse precision matrix is ideally a mixture of a point mass at zero and a continuous component \citep{wong_efficient_2003, carter_constructing_2011,talluri_bayesian_2014, banerjee_bayesian_2015}. However, since the normalizing constants in these mixture priors are intractable due to the positive definiteness constraint on the precision matrix, absolutely continuous priors have been proposed.  The Bayesian graphical lasso \citep{wang_bayesian_2012} has been developed as a Bayesian counterpart to the graphical lasso. However, its use of a double exponential prior, which does not have enough mass at zero, does not give a true Bayesian model for  sparsity.  Continuous shrinkage priors, such as the horseshoe \citep{carvalho_horseshoe_2010}, generalized double Pareto \citep{armagan_generalized_2013}, Dirichet-Laplace \citep{bhattacharya_dirichlet-laplace_2015}, and others have been proposed as better models of sparsity since these priors have infinite spikes at zero and heavy tails. 

Only a few results on the frequentist behavior of Bayesian methods for precision matrix estimation exist in the literature.  \cite{banerjee_posterior_2014} studied posterior convergence rates for a G-Wishart prior inducing a banding structure, but the true precision matrix need not have a banded structure.  \cite{banerjee_bayesian_2015} provided results on posterior contraction rates for the precision matrix under point mass spike-and-slab priors. 

Although GGMs are useful, the distributional assumption may fail to hold on some occasions.  A nonparametric extension of the normal distribution is the nonparanormal distribution in which the random variables $\bm{X} = (X_1,\ldots,X_d)$ are replaced by some transformed random variables $\bm{f}(\bm{X}) := (f_1(X_1),\ldots,f_d(X_d))$ and it is assumed that $\bm{f}(\bm{X})$ has a $d$-variate normal distribution $\N_d(\bm{\mu}, \bm{\Sigma}$) \citep{liu_nonparanormal:_2009}. In some situations, the logarithmic transform may be appropriate, but in general, the transformations $f_1,\ldots,f_d$ are hard to specify. It is, therefore, more sensible to let $f_1,\ldots,f_d$ be unspecified, and use a nonparametric technique for their estimation.  \cite{liu_nonparanormal:_2009} designed the nonparanormal graphical model, a two-step estimation process in which the functions $f_j$ were estimated first using a truncated empirical distribution function, and then the inverse covariance matrix $\boldsymbol\Omega = \boldsymbol\Sigma^{-1}$ was estimated using the graphical lasso applied to the transformed data. Although the approach in \cite{liu_nonparanormal:_2009} works well in many settings, their estimator for the transformation functions is based on the empirical distribution function, which leads to an unsmooth estimator. While the focus of this paper is on the nonparanormal graphical model, an alternative to the nonparanormal graphical model is the copula Gaussian graphical model \citep{pitt_efficient_2006, dobra_copula_2011, liu_high-dimensional_2012,mohammadi_bdgraph:_2017}  which avoids estimation of the transformation functions by using rank-based methods to transform the observed variables. 

Bayesian approaches can naturally blend the desired smoothness in the estimate by considering a prior on a function space that consists of smooth functions.  Gaussian process priors are the most commonly used priors on functions \citep{rasmussen_gaussian_2006,choudhuri_nonparametric_2007, van_der_vaart_bayesian_2007,lenk_bayesian_2017}.  Priors on function spaces have also been developed using a finite random series of certain basis functions like trigonometric polynomials, B-splines, or wavelets \citep{rivoirard_posterior_2012,  de_jonge_adaptive_2012, arbel_bayesian_2013,   shen_adaptive_2015}. 
We consider a Bayesian approach using a finite random series of B-splines prior on the underlying transformations.  We choose the B-splines basis over other possible choices  because B-splines can easily accommodate restrictions on functions, such as monotonicity and linear constraints, without compromising good approximation properties \citep{shen_adaptive_2015}. In our context, as the transformation functions $f_1,\ldots,f_d$ are increasing, imposing the monotonicity restriction through the prior is essential. This can be easily installed through a finite random series of B-splines by imposing the order restriction on the coefficients. By equipping the vector of the coefficients with a multivariate normal prior truncated to the cone of ordered coordinates, the order restriction can be imposed maintaining the conjugacy inherited from the original multivariate normal distribution. A simple Gibbs sampler is constructed in which first, a truncated normal prior on the transformation functions results in a truncated normal posterior distribution that is sampled using a Hamiltonian Monte Carlo technique \citep{pakman_exact_2014}.  Second, a Student t-spike-and-slab prior on the precision matrix of the transformed variables results in sampling the corresponding posterior distribution of the precision matrix and the edge matrix, which determines the absence or presence of an edge in the graphical model. The underlying graphical structure can then be constructed from the obtained edge matrix. 

The paper is organized as follows. In the next section, we state model assumptions of the Gaussian graphical model and the nonparanormal graphical model.  In addition, we specify the prior distributions for the underlying parameters.  In Section 3, we obtain the posterior distributions, describe the Gibbs sampling algorithm and the tuning procedure.  In Section 4, we provide a posterior consistency result for the priors under consideration. In Section 5, we present a simulation study.  In Section 6, we apply the method to a real data set and finally, we conclude with a discussion section.

\section{Model and Priors}
Let $\bm{X} = (X_1,\ldots,X_p)$ denote a random vector that is distributed as $p$-variate multivariate normal, $\N_p(\boldsymbol\mu, \boldsymbol\Sigma)$. The undirected graph $G = (V, E)$ that corresponds to this distribution consists of a vertex set $V$, which has $p$ elements for each component of $X$, and an edge set $E$ which consists of ordered pairs $(d, k)$ where $(d, k) \in E$ if there is an edge between $X_d$ and $X_k$.  The edge between $(d, k)$ is excluded from $E$ if and only if $X_d$ is independent of $X_k$ given all other variables. For multivariate normal distributions, the conditional independence holds if and only if $\boldsymbol\Sigma^{-1}_{d,k} = \boldsymbol\Omega_{d,k} = 0$; here for a matrix $\bm{A}$, $\bm{A}_{d,k}$ denotes its $(d,k)$th element.  

\begin{definition}
\label{def:nonparanormal}\rm 
A random vector $\bm{X} = (X_1,...,X_p)$ has a nonparanormal distribution if there exist smooth monotone functions $\{f_{d}: d=1,\ldots,p\}$ such that $\bm{Y} = \bm{f}(\bm{X}) \sim \N_p(\boldsymbol\mu, \boldsymbol\Sigma)$, where $\bm{f}(\bm{X}) = (f_{1}(X_1),\ldots,f_{p}(X_p))$. In this case we shall write $\bm{X}\sim \mathrm{NPN}(\bm{\mu},\bm{\Sigma},\bm{f})$. 
\end{definition}

By assuming that the transformed variables $\bm{f}(\bm{X})$ are distributed as normal, the conditional independence information in the nonparanormal model is completely contained in the parameter $\boldsymbol\Omega$, as in a parametric normal model. Since the transformation functions are one-to-one, the inherent dependency structure given by the graph for the observed variables is retained by the transformed variables. We note that any continuous random variable can be transformed into a normal variable by a strictly increasing transformation. However, testing for high-dimensional multivariate normality is not feasible, and hence testing for the nonparanormality assumption is not possible in high dimension, but clearly, the condition is a lot more general than multivariate normality. Instead of testing for nonparanormality, one may assess the efficacy of the assumption by looking at the effect of the transformations.  If the transformation functions are linear, then assuming multivariate normality should be adequate.  If the transformation functions are non-linear, then modeling through the nonparanormal distribution may be useful.

We put prior distributions on the unknown transformation functions through a random series based on B-splines. The coefficients are ordered to induce monotonicity, and the smoothness is controlled by the degree of the B-splines and the number of basis functions used in the expansion. Cubic splines, which are B-splines of degree 4, are used in this paper. The resulting posterior means of the coefficients give rise to a monotone smooth Bayes estimate of the underlying transformations. 

Thus the smooth monotone functions that we use to estimate the true transformation functions are assumed to be multivariate normal,
\begin{equation}
\label{spline model}
\bm{f}(\bm{X}) = \sum_{j=1}^J\boldsymbol\theta_jB_j(\bm{X}) \sim \N_p(\boldsymbol\mu, \boldsymbol\Omega^{-1}),
\end{equation}
where $\bm{f}$ is a $p$-vector of functions, $\bm{X}$ is an $n\times p$ matrix, and $\boldsymbol\theta_j$ is a $p$-vector; here $B_j(\cdot)$ are the B-spline basis functions, $\boldsymbol\theta_j$ are the associated coefficients in the expansion of the function, and $J$ is the number of B-spline basis functions used in the expansion. These transformed variables $\bm{f}(\bm{X})$ are subsequently used to estimate the sparse precision matrix and hence in structure learning.

In the next part, we discuss the prior on the coefficients in more detail.  

\begin{itemize}
\item \textbf{Prior on the B-spline coefficients}

First, we temporarily disregard the monotonicity issue and put a normal prior on the coefficients of the B-splines, $\boldsymbol\theta \sim \N_J(\boldsymbol\zeta, \sigma^2 \bm{I})$, where $\sigma^2$ is some positive constant, $\boldsymbol\zeta$ is some vector of constants, and $\bm{I}$ is the identity matrix.  A normal prior is convenient as it leads to conjugacy. However, apart from monotonicity of the transformations, we also need to address identifiability since unknown $\bm{\mu}$ and $\boldsymbol{\Sigma}$ allow flexibility in the location and the scale of the transformation so that the distribution of $\bm{f}(\bm{X})$ can be multivariate normal for many different choices of $\bm{f}$. The easiest way to address identifiability is to standardize the transformations by setting $\bm{\mu}=\bm{0}$ and the diagonal entries of $\boldsymbol{\Sigma}$ to $1$. However, then it will be more difficult to put a prior on sparse $\boldsymbol{\Omega}$ complying with the restriction on the diagonal entries of $\boldsymbol{\Sigma}$ because of the constraint $\boldsymbol{\Sigma}=\boldsymbol{\Omega}^{-1}$. Hence it is easier to keep $\bm{\mu}$ and $\bm{\Omega}$ free and impose restrictions on the locations and the scales of the transformation functions $f_d$, $d=1,\ldots,p$. There are different ways to impose constraints on the locations and scales of $f_d$.  One can impose some location and scale  restrictions on the corresponding B-spline coefficients, for instance, by making the mean $\bar{\theta}_d=J^{-1}\sum_{j=1}^J \theta_{dj}=0$ and the variance $\bar{\theta}_d=J^{-1}\sum_{j=1}^J (\theta_{dj}-\bar{\theta}_d)^2=1$. Then the prior distribution for $\bm{\theta}_d$, $d=1,\ldots,p$, will have to be conditioned on these  restrictions. The non-linearity of the variance restriction makes the prior less tractable. In order to obtain a conjugate normal prior, we instead consider the following
two linear constraints on the coefficients through function values of the transformations:
\begin{eqnarray}  
0 = f_d(1/2)&=&\sum_{j=1}^J\theta_{dj}B_j(1/2), \label{restriction1}\\
1 = f_d(3/4)-f_d(1/4)&=&\sum_{j=1}^J\theta_{dj}[B_j(3/4)-B_j(1/4)].\label{restriction2}
\end{eqnarray} 
It may be noted that, as only a few B-spline functions are non-zero at any given point, the restrictions \eqref{restriction1} and \eqref{restriction2} involve only a few $\theta_j$s. More specifically, as the degree of B-splines used in this paper is $4$, the first equation involves only $4$ coefficients and the second only $8$, no matter how large $J$ is.

The linear constraints can be written in matrix form as 
\begin{equation}
\label{restriction matrix form}
\bm{A}\boldsymbol\theta = \bm{c}, 
\end{equation}
where 
\begin{equation}
\label{expression A}
\bm{A}=\begin{bmatrix}
B_1(1/2) & B_2(1/2) & \cdots&B_J(1/2)  \\
B_1(3/4)-B_1(1/4) & B_2(3/4)-B_2(1/4) & \cdots & B_J(3/4)-B_J(1/4) \end{bmatrix}
\end{equation}
and $\bm{c}=(0,1)'$.

Using conditional normal distribution theory, the resulting prior on the coefficients $\boldsymbol\theta$ is 
$$ \boldsymbol\theta|\{\bm{A}\boldsymbol\theta = \bm{c}\} \sim \N_J(\boldsymbol\xi,\boldsymbol\Gamma),$$
where the prior mean and variance are
\begin{eqnarray}  
\boldsymbol\xi = \boldsymbol\zeta + \bm{A}'(\bm{A}\bm{A}')^{-1}(\bm{c}-\bm{A} \boldsymbol\zeta) 
\label{conditional mean}\\
\boldsymbol\Gamma = \sigma^{2}[\bm{I}-\bm{A}'(\bm{A}\bm{A}')^{-1}\bm{A}].
\label{conditional variance}
\end{eqnarray} 
However, the prior dispersion matrix $\boldsymbol\Gamma$ is singular due to the two linear constraints, resulting in a lack of Lebesgue density for the prior distribution on $\mathbb{R}^J$. Thus, we work with a dimension reduced coefficient vector by removing two coefficients to ensure that we have a Lebesgue density on $\mathbb{R}^{J-2}$ for the remaining components. 	Suppose we remove the last two coefficients.  Then, the reduced vector of basis coefficients is $\bar{\boldsymbol\theta}_d = [\theta_{d,1},\theta_{d,2},...,\theta_{d,J-2}]$.  Then we can solve for $\theta_{d, J-1}$ and $\theta_{d,J}$ using $\bm{A}\boldsymbol\theta = \bm{c}$ to obtain,
\begin{equation}
\label{expression two Bsplines}
    \begin{bmatrix}
  \theta_{d,J-1}\\
  \theta_{d,J}
\end{bmatrix} = 
\begin{bmatrix}
a_{d,1} & a_{d,2} & \cdots & a_{d,J-2}\\
b_{d,1} & b_{d,2} & \cdots & b_{d,J-2}
\end{bmatrix} \times \bar{\boldsymbol\theta}_d + \begin{bmatrix}
a_{d,0} \\
b_{d,0}
\end{bmatrix}
\end{equation}
where $a_{d,0},\ldots,a_{d,J-2}, b_{d,0},\ldots,b_{d,J-2}$ are the corresponding constants.  In matrix form, we have
\begin{equation}
\label{matrix expression two Bsplines}
    \begin{bmatrix}
  \theta_{d,J-1}\\
  \theta_{d,J}
\end{bmatrix} = \bm{W}_d\bar{\boldsymbol\theta}_d + \bm{q}_d, 
\end{equation} 
where $\bm{W}_d = \begin{bmatrix}
a_{d,1} & a_{d,2} & \cdots & a_{d,J-2}\\
b_{d,1} & b_{d,2} & \cdots & b_{d,J-2}
\end{bmatrix}$ and $\bm{q}_d = \begin{bmatrix}
a_{d,0} \\
b_{d,0}
\end{bmatrix}$.

Then the resulting prior for the coefficients for each predictor is,
\begin{equation}
\label{constrained prior}
\bar{\boldsymbol\theta}|\{\bm{A}\boldsymbol\theta = \bm{c}\} \sim \N_{J-2}(\bar{\boldsymbol\xi},\; \bar{\boldsymbol\Gamma}), 
\end{equation}
where the reduction is denoted with a bar. 

Finally, we impose the monotonicity constraint on the coefficients, which is equivalent with the series of inequalities $\theta_2- \theta_1 > 0,\ldots, \theta_J - \theta_{J - 1} > 0$ and expressed in matrix/vector form is $\bm{F}\boldsymbol\theta > \mathbf{0}$, where $\bm{F}$ is $(J - 1) \times J$,
\begin{equation}
\label{constraint full form}
\bm{F} = \begin{bmatrix} -1 & 1 & 0&\cdots &0&0\\
0 &-1&1& \cdots &0&0\\
\cdots \\
0&0&0& \cdots &-1&1
\end{bmatrix}.
\end{equation}

Due to the two linear constraints, the monotonicity constraint reduces to 
\begin{equation}
\label{constraint reduced form}
\bar{\bm{F}}\bar{\boldsymbol\theta} + \bar{\bm{g}} > \bm{0},
\end{equation}
where $\bar{\bm{F}}$ is the $(J-1) \times (J-2)$ matrix,
\begin{equation}
\label{matrix constraint reduced form}
\bar{\bm{F}} = \begin{bmatrix} -1 & 1 & 0&\cdots &0&0\\
0 &-1&1& \cdots &0&0\\
\cdots \\
0&0&0& \cdots &-1&1\\
a_1 & a_2 & a_3& \cdots &a_{J-3} & (a_{J-2}-1)\\
(b_1-a_2) & (b_2-a_2) & (b_3-a_3) &\cdots & (b_{J-3}-a_{J-3}) & (b_{J-2} - a_{J-2})
\end{bmatrix}
\end{equation}
 and $\bar{\bm{g}}$ is the constant $(J-2)$-vector,
 $\bar{\bm{g}} = (0, 0,0, \ldots, a_0, (b_0-a_0))'$.

The final prior on the coefficients is given by a truncated normal prior distribution
\begin{equation}
\label{final prior}
\bar{\boldsymbol\theta}|\{\bm{A}\boldsymbol\theta = \bm{c}\} \sim \mathrm{TN}_{J-2}(\bar{\boldsymbol\xi}, \bar{\boldsymbol\Gamma}, \mathcal{T}), 
\end{equation}
where $\mathcal{T} = \{\bar{\bm{\theta}}: \bar{\bm{F}}\bar{\boldsymbol\theta} + \bar{\bm{g}} >\mathbf{0}\}$, and 
the $\N_p(\bm{\mu},\bm{\Sigma})$-distribution restricted on a set $\mathcal{T}$ is denoted by $\mathrm{TN}_p(\bm{\mu}, \bm{\Sigma}, \mathcal{T})$. The conjugacy property of the prior distribution is preserved by the truncation.  
Instead of the simplifying example of solving for the last two coefficients, we use a more general method to reduce the dimension.  The Symbolic Math Toolbox in MATLAB is used to solve for any two coefficients in terms of the remaining coefficients.  In particular, for the first row of the linear constraints matrix $\bm{A}$ given by (2.5), we find the first column with a nonzero element.  Then, for the second row of the linear constraints matrix, we find the first column with a nonzero element that is not the same as the column selected from the first row.  We use the indices from those two columns to select the two coefficients that will be removed from the dimension in order to find $\bar{\boldsymbol\theta}, \bar{\bm{F}}$, and $\bar{\bm{g}}. $

Although any choice of $\bm{\zeta}$ is admissible, the prior can put a substantial probability of the truncation set $\mathcal{T} = \{\bar{\bm{\theta}}: \bar{\bm{F}}\bar{\boldsymbol\theta} + \bar{\bm{g}} >\mathbf{0}\}$ only when the original mean vector $\bm{\zeta}$ has increasing components. A simple choice of $\bm{\zeta}$ involving only two hyperparameters is given by  
\begin{equation}
\label{choice prior mean}
\zeta_j = \nu + \tau \Phi^{-1}\big(\frac{j-0.375}{J-0.75+1}\big), \; j= 1, \ldots J,
\end{equation}
where $\nu $ is a constant, $\tau$ is a positive constant, and $\Phi^{-1}$ is the inverse of the cumulative distribution function (i.e. the quantile function) of the standard normal distribution. The motivation for the choice comes from imagining that the prior distribution of each $\theta_j$ as $\N(\nu,\tau^2)$ before the ordering is imposed, and hence the expectations of the order statistics of $\N(\nu,\tau^2)$ may be considered as good choices for their means. The expression in \eqref{choice prior mean} gives a reasonable approximation to these expectations. Similar expressions $\Phi^{-1}(j/(J+1))$ appear for the score function of locally most powerful rank tests against normal alternatives (see \cite{hajek_theory_1999}). \cite{royston_algorithm_1982} described the expression $\Phi^{-1}(({j-0.375})/({J-0.75+1}))$, $j=1,\ldots,J$, as a more accurate approximation for the expected values of standard normal order statistics than the expression $\Phi^{-1}(j/(J+1))$ used in rank tests. 

\item \textbf{Prior on the mean}

For each predictor, we put an improper uniform prior $p(\bm{\mu})=\prod_{d=1}^p p_d(\mu_d)\propto 1$ on $\bm{\mu}$. 

\item \textbf{Prior on the precision matrix} 

We build on the techniques of \cite{wang_scaling_2015}, which use a normal spike-and-slab prior to estimate a sparse precision matrix, but replace the normal by a Student t-distribution spike-and-slab prior, following \cite{scheipl_spike-and-slab_2012}. Let $\tau_{d,k}^2$ be the slab variance and $c_0\tau_{d,k}^2$ be the spike variance.  The spike scale $c_0$ is assumed to be very small and given. Having a continuous spike instead of a point mass at zero is more convenient since it admits density; see \cite{wang_scaling_2015}. Unlike in \cite{wang_scaling_2015}, we estimate the sparse precision matrix by allowing the spike-and-slab variances and probability to be random with an inverse-gamma prior to lead to a Student t-distribution for the slabs. The diagonal entries of $\bm{\Omega}$ are given an exponential distribution with rate parameter $\lambda/2$ for some $\lambda>0$. We introduce a symmetric matrix of latent binary variables $\bm{L} = (\!(l_{d,k})\!)$ with binary entries to represent the edge matrix. The entries $l_{d,k}$, $d<k$, are assumed to be independent with $\pi$ denoting the probability of $1$, i.e. the probability of an edge. Let $\mathrm{N}(\cdot|\cdot,\cdot)$ and $\mathrm{Exp}(\cdot|\cdot)$ respectively stand for the densities of the normal and exponential distributions. Let $\bm{\eta} = (\tau_{d,k}^2,\pi , d<k,\lambda)$. Let $\mathcal{M}^{+}$ stand for the space of positive definite matrices and $v_{d,k}^2 = l_{d,k}\tau_{d,k}^2+c_0\tau_{d,k}^2 (1-l_{d,k})$.  The joint prior for $\boldsymbol\Omega= (\!(\omega_{d,k})\!)$ and $\bm{L}$ is then obtained as 
\begin{equation}
\label{omega_prior}
p(\boldsymbol\Omega,\bm{L}|\bm{\eta}) \propto \prod_{d<k}\textup{N}(\omega_{d,k}|0, v_{d,k}^2)
\prod_d\{\textup{Exp}(\omega_{d,d} |{\lambda}/{2})\}\prod_{d<k} \pi ^{l_{d,k}} (1-\pi )^{l_{d,k}} \mathbbm{1}_{\boldsymbol\Omega \in \mathcal{M}^{+}}. 
\end{equation}

The prior for $\bm{\eta}=(\tau_{d,k}^2,\pi , d<k,\lambda)$ are given by, independently of each other,  
\begin{equation}
\label{prior eta}
\pi  \sim \textup{Be}(1,10), \qquad 
\tau_{d,k}^2 \sim \mathrm{IG}(b_0,b_1),
\end{equation}
where Be stands for the beta distribution and IG for the inverse-gamma distribution.  The value of $\lambda$ controls the distribution of the diagonal elements of $\boldsymbol\Omega$.  We use $\lambda = 1$ under similar reasoning to \cite{wang_scaling_2015}, because it assigns a considerable probability to the region of reasonable values of the diagonal elements.  We set the shape parameters of the beta distribution to 1 and 10 to set the prior probability of sparsity to about 10\%.  See \cite{scheipl_spike-and-slab_2012} for more details regarding the spike-and-slab prior based on a mixture of inverse gamma distributions. 
\end{itemize}

\section{Posterior Computation}
The full posterior distribution is
\begin{eqnarray}
p(\boldsymbol\theta,\boldsymbol\Omega, \bm{L}, \boldsymbol\mu|\bm{X}) & \propto & 
(\det\boldsymbol\Omega)^{n/2}\exp\big(-\frac{1}{2}\sum_{i=1}^n (\boldsymbol\theta'\bm{B}(\bm{X}_i) - \boldsymbol\mu)'\boldsymbol\Omega(\boldsymbol\theta'\bm{B}(\bm{X}_i) - \boldsymbol\mu)\big) \nonumber \\
&& \times \prod_{d=1}^p p_d(\boldsymbol\theta_d)\prod_{d=1}^p p(\mu_d) \times  p(\boldsymbol\Omega, \bm{L})\mathbbm{1}_{\{\bm{F}\boldsymbol\theta > \bm{0} \}},
\label{full posterior}
\end{eqnarray}
where $\bm{B}(\bm{x})=(\!(B_j(x_d))\!)$, the prior on the B-spline coefficients is $ p_d(\boldsymbol\theta_d)$, the prior on the means is $p(\mu_d)$, and the joint prior on the sparse precision matrix and the edge matrix is $p(\boldsymbol\Omega, \bm{L})$. Here, the likelihood is constructed from the working assumption that  $\sum_{j=1}^J\boldsymbol\theta_jB_j(\bm{X})\sim \N_p(\boldsymbol\mu, \boldsymbol\Omega^{-1})$.

The joint posteriors are standard and so they are not derived.  They can be evaluated in the following Gibbs sampling algorithm.

\subsection{Gibbs Sampling Algorithm}

\begin{enumerate}
\item For every  $d = 1,\ldots, p$, sample the B-spline coefficients as follows.

\begin{enumerate}[(a)]
\item Since we can reduce the number of coefficients by two, the basis functions for these two coefficients can be represented as \[ \begin{bmatrix}
B_{J-1}(X_i) & B_J(X_i)
\end{bmatrix}
\begin{bmatrix}
  \theta_{d,J-1}\\
  \theta_{dJ}
\end{bmatrix} = \bm{B}^{*}\boldsymbol\theta_d^{*} = \bm{B}^{*}(\bm{W}_d\bar{\boldsymbol\theta}_d + \bm{q}_d) 
,\]
where the $*$ is used to denote the two-dimensional vectors $\bm{B}^{*}$ and $\boldsymbol\theta_d^{*}$. 

Setting $\bm{Y}_d =  (\sum_{j=1}^J\theta_{dj}B_j(X_{id}), d=1,\ldots,p,\, i = 1, \ldots, n)$, the joint posterior for the B-spline coefficients is a truncated normal, with density 
$$p(\bar{\boldsymbol\theta}_1,\ldots,\bar{\boldsymbol\theta}_p|\boldsymbol\Omega, \boldsymbol\mu, \bm{Y}) \propto (\det \boldsymbol \Omega)^{n/2}\exp\big(-\frac{1}{2}\sum_{i=1}^n(\bm{Y}_i - \boldsymbol\mu)'\boldsymbol\Omega(\bm{Y}_i - \boldsymbol\mu)\big)\times p(\bar{\boldsymbol\theta}_1) \cdots  p(\bar{\boldsymbol\theta}_p) $$
restricted on the region 
$\{\bar{\bm{F}}\bar{\boldsymbol\theta} + \bar{\bm{g}} > \mathbf{0} \}  $ to satisfy the monotonicity constraint. 

However, this truncated multivariate normal distribution is $p\times (J-2)$ dimensional, so we sample it using the following conditional normals in a Markov chain,
\begin{multline*}
p(\bar{\boldsymbol\theta}_d|\bm{Y}, \bar{\boldsymbol\theta}_{\{1,\ldots, p\}\setminus d}, \boldsymbol\mu, \boldsymbol\Omega) \propto  \exp  \Big [-\frac{1}{2} \bar{\boldsymbol\theta}_d'\big \{ \frac{1}{\lambda_d^2}\sum_{i=1}^n(\bar{\bm{B}} + \bm{B}^{*}\bm{W}_d)'(\bar{\bm{B}} + \bm{B}^{*}\bm{W}_d) + \bar{\boldsymbol\Gamma}^{-1} \big\}\bar{\boldsymbol\theta}_d \\
+ \big \{ \bar{\boldsymbol\xi} \bar{\boldsymbol\Gamma}^{-1}  - \frac{1}{\lambda_d^2}\sum_{i=1}^n (\bm{B}^{*}\bm{q}_d -\delta_{d,i})'(\bar{\bm{B}} + \bm{B}^{*}\bm{W}_d)\big \} \bar{\boldsymbol\theta}_d 
\Big ]\mathbbm{1}_{\{\bar{\bm{F}}_d\bar{\boldsymbol\theta}_d + \bar{\bm{g}}_d > \mathbf{0} \}},
\end{multline*}
where using the conditional normal theory, 
$$\delta_{d,i}= \mu_d + \sum_{e \in (1:p\setminus d)} (-\frac{\boldsymbol\omega_{d,e}}{\omega_{d,d}})(Y_{i,e} - \mu_e)  $$ 
and $\lambda_d^2 = {1}/{\omega_{d,d}} $. 

Samples from the truncated conditional normal posterior distributions for the B-spline coefficients are obtained using the exact Hamiltonian Monte Carlo algorithm (exact HMC) \citep{pakman_exact_2014}.  
Each iteration of the exact HMC results in a transition kernel which leaves the target distribution invariant and the Metropolis acceptance probability equal to 1. The exact HMC within Gibbs is like Metropolis within Gibbs and hence is a valid algorithm to sample from the joint density. 

\end{enumerate}
\item Obtain the centered transformed variables: 
\begin{enumerate}[(a)]
\item Compute $Y_{id} = \sum_{j=1}^J\theta_{dj}B_j(X_{id})$; 
\item  Sample $\boldsymbol\mu|(\bm{Y}, \boldsymbol\Omega) \sim \N_p(\bar{\bm{Y}}, \frac{1}{n}\boldsymbol\Omega^{-1})$; 
\item Find $Z_{id}  = \sum_{j=1}^J\theta_{dj}B_j(X_{id}) -\mu_d  = Y_{id} - \mu_d$. 
\end{enumerate} 

\item The posterior density of $\bm{\Omega}$ given $\bm{L}$ is
$$p(\boldsymbol\Omega | \bm{Z}, \bm{L},\boldsymbol\tau^2,\lambda) \propto (\det \boldsymbol\Omega)^{n/2}\exp\big\{-\frac{1}{2}\textup{tr}(\bm{S}\boldsymbol\Omega)\big\}\prod_{d<k}\exp\big(-\frac{\omega_{d,k}^2}{2v_{d,k}^2}\big)\prod_{d=1}^p\exp\big(-\frac{\lambda}{2}\omega_{d,d}\big),$$
where $ \bm{S} = \bm{Z}'\bm{Z}$.

For every $d = 1,\ldots,p$, sample each column vector of $\boldsymbol\Omega$ and $\bm{L}$ using the following partitions as described in \cite{wang_scaling_2015}: 
\begin{itemize}
\item 
Denote $\bm{V} = (\!(v_{d,k}^2)\!)$ to be the $p \times p$ symmetric matrix with zeros in the diagonal and $ (v_{d,k}^2 = l_{d,k}\tau_{d,k}^2+c_0\tau_{d,k}^2 (1-l_{d,k}): {d < k}) $ in the upper diagonal entries.  Similarly, denote $\bm{T}=  (\!(\tau_{d,k}^2)\!)$ and $\boldsymbol\Pi=  (\!(\pi_{d,k})\!)$ to be $p \times p$ symmetric matrices with zeros in the diagonal and $ (\tau_{d,k}^2: {d < k}) $ and $ (\pi_{d,k}: {d < k}) $ in the upper diagonal entries, respectively. 
\item 
Without loss of generality, partition 
  $\boldsymbol\Omega, \bm{S}, \bm{L}, \bm{V}, \bm{T}$, and $\boldsymbol\Pi$ by focusing on the last column and row: 
\[
\boldsymbol\Omega =  
\begin{bmatrix}
\boldsymbol\Omega_{11} & \boldsymbol\omega_{12}\\
\boldsymbol\omega_{12}' & \omega_{22}
\end{bmatrix},
\qquad 
\bm{S} = 
\begin{bmatrix}
\bm{S}_{11} & \bm{s}_{12}\\
\bm{s}_{12}' & s_{22}
\end{bmatrix},
\qquad 
\bm{L} =  
\begin{bmatrix}
\bm{L}_{11} & \bm{l}_{12}\\
\bm{l}_{12}' & l_{22}
\end{bmatrix},
\]
\[\bm{V} = 
\begin{bmatrix}
\bm{V}_{11} & \bm{v}_{12}\\
 \bm{v}_{12}' & 0
\end{bmatrix},
\qquad
\bm{T} = 
\begin{bmatrix}
\bm{T}_{11} & \boldsymbol\tau_{12}\\
 \boldsymbol\tau_{12}' & 0
\end{bmatrix}, 
\qquad
\boldsymbol\Pi = 
\begin{bmatrix}
 \boldsymbol\Pi_{11} & \boldsymbol\pi_{12}\\
 \boldsymbol\pi_{12}' & 0
\end{bmatrix}.
\]
\item To sample a column vector of $\boldsymbol\Omega$, use the following change of variables: $$(\boldsymbol\omega_{12},\omega_{22}) \mapsto (\bm{u} = \boldsymbol\omega_{12},\, v = \omega_{22} - \boldsymbol\omega_{12}'\boldsymbol\Omega_{11}^{-1}\boldsymbol\omega_{12}).$$  Then the full conditionals are given by 
$$(\bm{u}|\cdot)\sim \N(-\bm{C}\bm{s}_{12}, \bm{C}),\,  (v|*)\sim \textup{Ga}\big(\frac{n}{2}+1,\frac{s_{22}+\lambda}{2}\big),  $$
where $\bm{C} = \{(s_{22} + \lambda)\boldsymbol\Omega_{11}^{-1} + \textup{diag}(\textbf{v}_{12}^{-1}) \}^{-1}$, and Ga stands for the gamma distribution.

\item To sample the corresponding off-diagonal column vector of the edge-inclusion indexes  ${l}_{dk}$, $d,k=1,\ldots,p$, $d<k$, since the $l_{d,k}$ are independent Bernoulli, we sample according to the probability 
$$\textup{P}({l}_{dk} = 1|\cdot) = \frac{\phi(\omega_{dk}|0, \tau_{dk}^2)\pi_{dk}}{\phi(\omega_{dk}|0, \tau_{dk}^2)\pi_{dk} + \phi(\omega_{dk}|0, c_0\tau_{dk}^2)(1-\pi_{dk})}.$$
where $\phi$ stands for the normal density function. 
\item Update $\tau_{dk}^2$, $d,k=1,\ldots,p$, $d<k$, based on the off-diagonal column vectors $\omega_{dk}$ and  ${l}_{dk}$, using the relations 
$$(\tau_{dk}^2 |\cdot) \sim \mathrm{IG}\big(b_0+\frac{1}{2},b_1+\frac{\omega_{dk}^2}{2}({l}_{dk} + \frac{1-{l}_{dk}}{c_0})\big).$$

\item Update $\pi $, $d,k=1,\ldots,p$, $d<k$, based on the off-diagonal entry ${l}_{dk}$,
$$(\pi |\cdot) \sim \textup{Be}(1+ \sum_{d<k}\mathbbm{1}\{{l}_{dk}=1\}, 10+ \sum_{d<k}\mathbbm{1}\{{l}_{dk} = 0\}). $$
\end{itemize}
\end{enumerate}
These steps are repeated until convergence.

\subsection{Choice of Prior Parameters}
\label{ChoiceParameters}
We use a model selection criterion to determine the optimal number of basis functions pre-MCMC.   Sampling methods that involve putting a prior on the number of basis functions, such as reversible jump Monte Carlo, are computationally complicated.  We calculate the Akaike Information Criterion (AIC) for different numbers of basis functions and choose the number of basis functions that correspond to the lowest AIC.  The AIC is determined as the minimum of two times the negative log-likelihood $-2 l(\boldsymbol\theta_d)$, plus the number of parameters in the model, with respect to the basis coefficients subject to the linear and monotonicity constraints. The AIC is preferred here as the true transform does not belong to the set of splines and hence the correct model selection is not the goal, but minimizing the estimated estimation error is, which is provided by the model with the lowest AIC. The lowest AIC is found between a grid of four and 100 basis functions by doing a search in which the lowest AIC is chosen when the next ten values are larger than the current value in the search, since the AIC should approximately be a U-shaped curve due to the trade-off between accuracy and complexity.  Then for each predictor, $d = 1,\ldots,p$, and for the number of basis functions, $J$,
\begin{equation}
-2l(\boldsymbol\theta_d) =  n\log \sigma_d^2 + \frac{1}{\sigma_d^2}\sum_{i=1}^n \Big(\sum_{j=1}^J\theta_{dj}B_j(X_{id})-\mu_d \Big)^2. 
\end{equation}

After plugging in the maximum likelihood estimators (MLEs) of $\mu_d$ and $\sigma_d$ and making the substitution $Z_{id} = B_j(X_{id}) - n^{-1}\sum_{m=1}^nB_j(X_{md})$, minimizing the $-2l(\boldsymbol\theta_d)$ results in the following problem,
\begin{equation}
\underset{\boldsymbol\theta_d}{\text{minimize }} 
 n\log (\boldsymbol\theta_d'\bm{Z}'\bm{Z}\boldsymbol\theta_d), \quad\text{subject to }
 \bm{F}\boldsymbol\theta_d > \mathbf{0}, \; 
\bm{A}\boldsymbol\theta_d = \bm{c}. 
\end{equation}
This problem can be equivalently solved using the quadratic
programming function in MATLAB Optimization Toolbox:

\begin{equation}
\underset{\boldsymbol\theta_d}{\text{minimize }} 
\frac{1}{2}\boldsymbol\theta_d'\bm{Z}'\bm{Z}\boldsymbol\theta_d, \quad\text{subject to }
 \bm{F}\boldsymbol\theta_d > \mathbf{0}, \; 
\bm{A}\boldsymbol\theta_d = \bm{c}. 
\end{equation}

For numerical stability, the monotonicity constraint was changed to $\textbf{F}\boldsymbol\theta_d \geq 10^{-4}$. 
Finally, after plugging in the solution of the quadratic programming problem $\hat{\boldsymbol\theta}_d$, the final number of basis functions is chosen by selecting the number $J$ that minimizes the AIC
$$\textup{AIC} = -2l(\hat{\boldsymbol\theta}_d) + 2J =  n\log  (\hat{\boldsymbol\theta}_d'\bm{Z}'\bm{Z}\hat{\boldsymbol\theta}_d)  + 2J.$$

There is some dependence on the choice of hyperparameters.  We use a model selection criterion to determine the hyperparameters, $b_0$ and $b_1$, for inverse gamma distributions for $(\!(\tau^2_{dk})\!)$ and to determine the constant value for the spike scale, $c_0$, after the MCMC sampling.  Inspired by \cite{dahl_maximum_2005,dahl_covariance_2008}, we solve a convex optimization problem in order to use the Bayesian Information Criterion (BIC).  First, we find the Bayes estimate of the inverse covariance matrix, $\hat{\boldsymbol\Omega}_{\mathrm{Bayes}}$. The Bayes estimate is defined as $\hat{\boldsymbol\Omega} = \mathrm{E}(\boldsymbol\Omega | \bm{Z})$.  We find the average of the transformed variables, $\bar{\bm{Z}} = M^{-1}\sum_{m = 1}^M\bm{Z_m}$, where $\bm{Z}_m$, $m = 1, \ldots, M$, are obtained from the MCMC output.  Then, using the sum of squares matrix, $\bm{S} =\bar{\bm{Z}}'\bar{\bm{Z}}$, we solve the following to obtain the maximum likelihood estimate of the inverse covariance matrix, $\hat{\boldsymbol\Omega}_{\mathrm{MLE}}$:
\begin{equation}
\underset{\boldsymbol\Omega}{\text{minimize }} 
-n\log\det \boldsymbol\Omega +\textup{tr}(\boldsymbol\Omega \bm{S}), \quad 
 \text{subject to }
\mathcal{C}(\hat{\boldsymbol\Omega}),
\end{equation}
where $\mathcal{C}$ represents the elements of $\hat{\boldsymbol\Omega}$ that are zero and nonzero, and they are determined by the zeros of the estimated edge matrix from the MCMC.  The estimated edge matrix from the MCMC sampler will be described in more detail in Section 5.  This constrained optimization problem was implemented as an unconstrained optimization problem, as described in \cite{dahl_maximum_2005,dahl_covariance_2008}.

Finally, we calculate 
$\textup{BIC} = -2l(\hat{\boldsymbol\Omega}_{\mathrm{MLE}}) + k\log n  $, 
where $k = \#\mathcal{C}(\hat{\boldsymbol\Omega})$, the sum of the number of diagonal elements and the number of edges in the estimated edge matrix, and $-l(\hat{\boldsymbol\Omega}_{\mathrm{MLE}}) = -n\log \det \hat{\boldsymbol\Omega}_{\mathrm{MLE}} + \textup{tr}(\hat{\boldsymbol\Omega}_{\mathrm{MLE}}\bm{S})$. 

We select the combination of hyperparameters, $b_0$, $b_1$, $c_0$, that results in the smallest BIC.  

\section{Posterior Consistency}

Posterior consistency is a fundamental way of validating a Bayesian method using a frequentist yardstick in the large sample setting, and is of interest to both frequentists and Bayesians; for a thorough account of posterior consistency, see \cite{ghosal_fundamentals_2017}. In Gaussian graphical models, using point mass spike-and-slab priors, \cite{banerjee_bayesian_2015} showed that the posterior for $\bm{\Omega}$ is consistent in the high-dimensional setting provided that $(p+s)(\log p)/n\to 0$, where $s$ stands for the number of non-zero off-diagonal entries of the true $\bm{\Omega}$. With a slight modification of the arguments, it follows that the result extends to continuous spike-and-slab priors provided that the spike scale $c_0$ is sufficiently small with increasing $p$. In the nonparanormal model, the main complicating factor comes from the unknown transformations $f_1,\ldots,f_p$, since the rest will then be as in a Gaussian graphical model. Below we argue that these transformations may be estimated consistently in an appropriate sense. 

We study the posterior distributions for each transformation $f_d$ separately, which can be learned from the marginal likelihood for each component. Thus the problem of posterior consistency for $f_d$ can be generically described as follows. For brevity, we drop the index $d$. Consider the model $Y = f(X) \sim \N(\mu, \sigma^2)$, where $f$ is a continuously differentiable, strictly monotone increasing transformation from $(0,1)$ to $\mathbb{R}$. Clearly, this model is not identifiable and hence consistent estimation is not possible in the usual sense. Identifiability can be ensured by setting $\mu=0$ and $\sigma=1$, but the procedure followed in this paper instead puts constraints on $f$: $f(1/2)=0$ and $f(3/4)-f(1/4)=1$. We shall show that the posterior for $f$ is consistent under this set of constraints. 

As the function $f$ is necessarily unbounded near $0$ and $1$ to ensure that $f(X)$ is normally distributed, which is a distribution with unbounded support, it is clear that uniform posterior consistency for $f$ is not possible. We shall, therefore, consider the notion of uniform convergence on a compact subset of $(0,1)$: for a fixed $\delta>0$, the pseudo-metric to consider is $d(f_1,f_2)=\sup\{|f_1(x)-f_2(x)|: \delta\le x\le 1-\delta\}$. Even then, the usual posterior distribution may be highly impacted by observations near $0$ or $1$, so we actually study a modified posterior distribution, based on observations falling within the given fixed compact subset $[\delta,1-\delta]$ of $(0,1)$, with $\delta<1/4$, to be described below.

Let $f_0$ be the true transformation function, which is assumed to be continuously differentiable and strictly monotone increasing and complying with the constraints $f_0(1/2)=0$ and $f_0(3/4)-f_0(1/4)=1$. Let $\mu_0$ and $\sigma_0>0$ be respectively the true values of $\mu$ and $\sigma$. Note that the cumulative distribution function (c.d.f.) of $X$ is given by $F(x)=\mathrm{P}(X\le x)=\mathrm{P}(f(X)\le f(x))=\Phi((f(x)-\mu)/\sigma)$ and the corresponding true c.d.f. is $F_0(x)=\Phi((f_0(x)-\mu_0)/\sigma_0)$, where $\Phi$ stands for the c.d.f. of the standard normal distribution.  

Consider $n$ i.i.d. observations $X_1,\ldots,X_n$ from the true distribution. 
Let $n^*$ be the number of observations falling in $[\delta,1-\delta]$, $n^*_-$ the number of observations falling below $\delta$ and $n^*_+$ the number of observations falling above $1-\delta$. Let $X_1^*,\ldots,X^*_{n^*}$ be the observations falling in $[\delta,1-\delta]$. The posterior consistency is based on the posterior given these complying observations $X_1^*,\ldots,X^*_{n^*}$, and the counts $(n^*_-,n^*_+)$. 

Observe that $\pi^-:=\mathrm{P}(X<\delta)=F(\delta)$ and $\pi^+:=\mathrm{P}(X>1-\delta)=1-F(1-\delta)$. Then $n^*_-\sim \mathrm{Bin}(n,\pi^-)$ and $n^*_+\sim \mathrm{Bin}(n,\pi^+)$. Let $F^*$ stand for the c.d.f. of $X^*$, let $F_0^*$ stand for its true value and let $(\pi^-_0,\pi^+_0)$ be the true value of $(\pi^-,\pi^+)$. Then we have the identity 
\begin{equation}
\label{eq:truncation}
F(x)=\pi^-+(1-\pi^+-\pi^-)F^*(x), \quad F_0(x)=\pi^-_0+(1-\pi^+_0-\pi^-_0)F^*_0(x)
\end{equation}
for all $x\in [\delta,1-\delta]$. 

Thus we have 
\begin{equation}
\label{transformation and cdf}
f(x)=\mu+\sigma \Phi^{-1}(F(x)), \quad f_0(x)=\mu_0+\sigma_0 \Phi^{-1}(F_0(x)).
\end{equation}

We note that the posterior distributions of the quantities $\pi^-$ and $\pi^+$ can be obtained based on the counts $n^*_-$ and $n^*_+$ respectively. In particular, using a Dirichlet prior on the probability vector $(\pi^-,\pi^+,1-\pi^--\pi^+)$, we have consistency for the posterior distribution of $(\pi^-,\pi^+)$ at $(\pi^-_0,\pi^+_0)$. We shall assume that the posterior distribution of $(\pi^-,\pi^+)$ is consistent. Note that the truncated  observations alone do not lead to a posterior distribution for $(\pi^-,\pi^+)$. 

The modification in the posterior distribution of $\mu$, $\sigma$ and $f$ that we consider can be described as follows. Using the given prior on $(\mu,\sigma,f)$ and the truncated observations $X^*_1,\ldots,X^*_{n^*}$, we obtain the induced posterior distribution of $F^*$, while we obtain the posterior distribution on $(\pi^-,\pi^+)$ directly conditioning on $(n^*_-,n^*_+)$. Then the posterior distribution of $\{F(x): x\in [\delta,1-\delta]\}$ is induced from \eqref{eq:truncation}. Finally, the modified posterior distribution of $(\mu,\sigma,f)$ is induced from the relations 
\begin{equation}
\label{mu sigma}
\sigma=1/(\Phi^{-1}(F(3/4))-\Phi^{-1}(F(1/4))), \quad \mu=-\frac{\Phi^{-1}(F(1/2))}{\Phi^{-1}(F(3/4))-\Phi^{-1}(F(1/4))}, 
\end{equation}
and \eqref{transformation and cdf} in view of the restrictions $f(1/2)=0$ and $f(3/4)-f(1/4)=1$. The corresponding true values satisfy the analogous relations 
\begin{equation}
\label{mu sigma true}
\sigma_0=1/(\Phi^{-1}(F_0(3/4))-\Phi^{-1}(F_0(1/4))), \quad \mu_0=-\frac{\Phi^{-1}(F_0(1/2))}{\Phi^{-1}(F_0(3/4))-\Phi^{-1}(F_0(1/4))}. 
\end{equation}

The following theorem on posterior consistency refers to this modified posterior distribution rather than the original posterior distribution of $(\mu,\sigma,f)$.  The proof can be found in the Supplementary Material.

\begin{thm} 
\label{thm:posterior consistency}
In the above setting let the prior on $\mu$ and $\sigma$ contain $\mu_0$ and $\sigma_0$ in its support and independently, the prior $\Pi$ for $f$ satisfies the condition that 
\begin{equation}
\label{eq:condition}
\Pi(f: d(f,f_0)<\epsilon, d(f',f_0')<\epsilon)>0 \mbox { for every }\epsilon>0.
\end{equation}
Then for any $\epsilon>0$, 
\begin{equation}
\label{eq:consistency}
\Pi(|\mu-\mu_0|<\epsilon, |\sigma-\sigma_0|<\epsilon, d(f,f_0)<\epsilon|X_1^*,\ldots,X_{n^*}^*, \, n^*_-,n^*_+)\to 1 \mbox{ a.s.}
\end{equation}
\end{thm}

The condition on the prior for the transformation $f$ is satisfied by the truncated normal prior described in Section~2, and hence the transformation  $f$ (as well as the mean and variance parameters $\mu$ and $\sigma^2$) are consistently estimated by the posterior, as shown in the following corollary. 

\begin{corollary}
Let the prior on $f$ be described by $f=\sum_{j=1}^J \theta_j B_j$, where the prior for $J$ has infinite support and $\bm{\theta}=(\theta_1,\ldots,\theta_J)$ is given a truncated normal prior as described in Section~2. Then for any $\epsilon>0$, $\Pi(f: d(f,f_0)<\epsilon, d(f',f_0')<\epsilon)>0$ and hence \eqref{eq:consistency} holds. 
\end{corollary}

\section{Simulation}

We conduct a simulation study to assess the performance of the Bayesian approach to graphical structure learning in nonparanormal graphical models.  This method will be referred to as `Spike Slab' in the results.  The unobserved random variables, $Y_1,\ldots,Y_p$, are simulated from a multivariate normal distribution such that $Y_{i1},\ldots,Y_{ip} \iidsim \N_p(\boldsymbol\mu, \boldsymbol\Omega^{-1})$ for $i = 1, \ldots, n$.  The means $\boldsymbol\mu$ are selected from an equally spaced grid between 1 and 2 with length $p$.  We consider nine different combinations of $n, p,$ and sparsity for $\boldsymbol\Omega$:

\begin{itemize}
\item $p=25$, $n=50$, sparsity = $10\%$ non-zero entries in the off-diagonals
\item $p=50$, $n=150$, sparsity = $5\%$ non-zero entries in the off-diagonals
\item $p=100$, $n=500$, sparsity = $2\%$ non-zero entries in the off-diagonals
\item $p=25$, $n=50$, AR(1) model
\item $p=50$, $n=150$, AR(1) model
\item $p=100$, $n=500$, AR(1) model
\item $p=25$, $n=50$, circle model
\item $p=50$, $n=150$, circle model
\item $p=100$, $n=500$, circle model
\end{itemize}
 where the circle model and the AR(1) model are described by the relations 
\begin{itemize} 
\item Circle model: $\omega_{ii} = 2,\; \omega_{i, i-1} = \omega_{i-1,i} = 1$, and $\omega_{1,p}=\omega_{p,1} = 0.9$
\item AR(1) model: $\omega_{11} = \omega_{pp} = 1.9608, \; \omega_{ii} = 2.9216$ and $\omega_{i, i-1} = \omega_{i-1,i} =  -1.3725$.
\end{itemize}

The sparsity levels for $\boldsymbol\Omega$ are computed using lower triangular matrices that have diagonal entries that are Gaussian
distributed with $\mu_{\textup{diag}} = 1$ and $\sigma_{\textup{diag}} = 0.1$, and 
non-zero off-diagonal entries that are Gaussian distributed
with $\mu_{\setminus \textup{diag}} = 0$ and $\sigma_{\setminus \textup{diag}} = 1$.  Since these are lower triangular matrices, we are ensured to have positive definite matrices.

 The hyperparameters for the prior are chosen to be $\nu = 1, \tau = 0.5,$ and $\sigma^2 = 1$.  The observed variables $\bm{X} = (X_1, \ldots,X_p)$ are constructed from the simulated unobserved variables $Y_1, \ldots, Y_p$.  The functions used to construct the observed variables are four c.d.f.s and the power function evaluated at the simulated unobserved variables $Y_1, \ldots, Y_p$.  The four c.d.f.s are:  normal, logistic, extreme value, and stable.  The power function is $X_d = [\Phi(Y_d)]^{1/m}$, $d=1,\ldots,p$, where $m$ is an integer between 1 and 5.  We could choose any value for the parameters, but for computational ease, we use the maximum likelihood estimates of the parameters with the {\tt mle} function in MATLAB.  Any values for the parameters could be chosen for the c.d.f.s.  We choose the values of the parameters for each of the c.d.f.s to be the maximum likelihood estimates  for the parameters of the corresponding distributions (normal, logistic, extreme value, and stable), using the variables $Y_1, \ldots, Y_p$.

The initial B-spline coefficient values for the exact HMC algorithm are constructed as follows.  First, pretending that the data are already normal, we start with the identity function $f(X_d) = X_d$, where $X_d$ is uniform and  $\Phi^{-1}(X_d)$ is normal so that $f=\Phi^{-1}$.  Then in the model, with the pretension that the transformation is a linear combination of B-spline basis functions,  $\Phi^{-1}(X_d) = \sum_{j=1}^J\theta_jB_j(X_d) \sim \N(\mu_d, \sigma_d^2)$. Multiplying both sides by $B_k(X_d)$ and integrating, we have $\int_0^1  B_k(X_d) \Phi^{-1}(X_d)dX_d = \sum_{j=1}^J\theta_j\int_0^1B_j(X_d)B_k(X_d)dX_d$. Making the substitution $Z_d = \Phi^{-1}(X_d)$, so that $\Phi(Z_d) = X_d$, leads to the relation  
$$\int_{-\infty}^{\infty}B_k(\Phi(Z_d))Z_d \phi(Z_d)dZ_d = \sum_{j=1}^J\theta_j\int_{-\infty}^{\infty}B_j(\Phi(Z_d))B_k(\Phi(Z_d))\phi(Z_d)dZ_d.$$  

Since these functions in the integral are functions of normal probability densities, Gauss-Hermite quadrature is used to estimate the left and right-hand sides.  The number of points used is 20.  Then setting the approximation for the left-hand side,  $\int_{-\infty}^{\infty}B_k(\Phi(Z_d))Z_d \phi(Z_d)dZ_d$, equal to $\bm{b}$, and setting the approximation for the right-hand side, $\int_{-\infty}^{\infty}B_j(\Phi(Z_d))B_k(\Phi(Z_d))\phi(Z_d)dZ_d$ equal to $\bm{E}$, we have the linear equation $\bm{b} = \bm{E}\boldsymbol\theta$.  Using the quadratic
programming function in the MATLAB Optimization Toolbox, we solve for $\boldsymbol\theta$ for each predictor
\begin{equation}
\underset{\theta}{\text{minimize }} 
\frac{1}{2} \boldsymbol\theta'\bm{E}'\bm{E}\boldsymbol\theta - \bm{b}'\bm{E}\boldsymbol\theta, 
 \text{ subject to }
 \bm{F}\boldsymbol\theta > \mathbf{0}, \; 
\bm{A}\boldsymbol\theta = \bm{c}. 
\end{equation}
For numerical stability, the monotonicity constraint is changed to $\bm{F}\boldsymbol\theta \geq 10^{-4}$.

After finding the initial coefficients $\boldsymbol\theta_d$, we construct the initial values for $Y_d =  \sum_{j=1}^J\theta_{dj}B_j(X_{d})$ using the observed variables.  These initial values for $Y$ are used to find initial values for $\boldsymbol\Sigma, \boldsymbol\mu$, and $\boldsymbol\Omega$ for the algorithm, where $\boldsymbol\Sigma_{\textup{initial}} = \textup{cov}(\bm{Y}), \boldsymbol\mu_{\textup{initial}} = \bar{\bm{Y}}$, and $\boldsymbol\Omega_{\textup{initial}} = \boldsymbol\Sigma_{\textup{initial}}^{-1}$.

We consider four combinations of the hyperparameter settings for the spike-and-slab algorithm with $c_0 = \{0.02, 0.005\}$ and $(b_0,b_1) = \{(1,1), (10,30)\}$.  We select the values for $b_0$ and $b_1$ because they cover reasonable shapes of the prior distribution for the slab variance and the $c_0$, assumed to be small, covers two different orders of magnitude.  Other choices could be used for the spike-and-slab algorithm.  The model selection criterion described in Section \ref{ChoiceParameters} chooses the combination of hyperparameters that yields the lowest BIC, and that combination of hyperparameters is used to obtain the final estimates of the precision matrix and edge matrix.  The spike-and-slab algorithm is implemented in MATLAB by modifying the code provided by \cite{wang_scaling_2015}.  The exact HMC algorithm is implemented in MATLAB using the code provided by the authors \citep{pakman_exact_2014}. 

The nonparanormal method of \cite{liu_nonparanormal:_2009} is implemented using the R package {\tt huge}  \citep{zhao_huge:_2015}.  The graphical lasso method is selected for the graph estimation and by default, the screening method selected is the lossless screening method.  Three regularization selection methods are used to select the graphical model: the Stability Approach for Regularization Selection (StARS) \citep{liu_stability_2010}, a modified Rotation Information Criterion (RIC) \citep{lysen_permuted_2009}, and the Extended Bayesian Information Criterion (EBIC) \citep{foygel_extended_2010}.  The default parameters in the {\tt huge} package are used for each selection method.  The documentation for the {\tt huge} package mentions an alternative threshold of 0.05 for the StARS method, but the results are not sensitive to the default choice of 0.1 or 0.05, so the default threshold of 0.1 is used.  As in \cite{liu_nonparanormal:_2009}, the number of regularization parameters used is 50 and they were selected among an evenly spaced grid in the interval [0.16,1.2].  

 A Bayesian copula graphical model \citep{mohammadi_bayesian_2017} is implemented using the {\tt R} package, {\tt BDgraph} \citep{mohammadi_bdgraph:_2017, mohammadi_bdgraph:_2019}. This method will be referred to as, `Bayesian Copula'.  Posterior graph selection is done using Bayesian model averaging, the default option in the package, in which it selects the graph with links for which their
estimated posterior probabilities are greater than 0.5.

We run 100 replications for each of the nine combinations and assess structure learning for each replication.  We collect $10000$ MCMC samples for inference after discarding a burn-in of $5000$.  Thinning is not applied.  For each replication, we determine the final hyperparameter setting for the Spike Slab method by choosing, out of the four hyperparameter settings, the one that yields the lowest value of the BIC.  Finally, the selected hyperparameter setting is used to find the Bayesian estimates of the precision and edge matrices and are used to learn the graphical structure.

To assess the performance of the graphical structure learning, specificity (SP), sensitivity (SE), and Matthews Correlation Coefficient (MCC) are computed.  These metrics have been previously used for assessing the accuracy of classification procedures \citep{baldi_assessing_2000}.  They are defined as follows:
\begin{align*}
\textup{Specificity} = \frac{\textup{TN}}{\textup{TN} + \textup{FP}}, \qquad \textup{Sensitivity} = \frac{\textup{TP}}{\textup{TP} + \textup{FN}},\\ 
\\
\textup{MCC} = \frac{\textup{TP} \times \textup{TN} - \textup{FP} \times \textup{FN}}{\sqrt{(\textup{TP} + \textup{FP})(\textup{TP} + \textup{FN})(\textup{TN} + \textup{FP})(\textup{TN} + \textup{FN})}},
\end{align*}
where TP is the number of true positives, TN is the number of true negatives, FP is the number of false positives, and FN is the number of false negatives.  True positives mean that edges that are included in the estimate are also present in the true model, true negatives mean that edges that are not included in the estimate are also not included in the true model, false positives mean that there are edges included in the estimate that are not present in the true model and false negatives mean that there are edges that are not included in the estimate that are present in the true model. The MCC is regarded as an overall measure of classification.  The higher the values are for all three metrics, the better is the classification. 

The median probability model \citep{berger_optimal_2004}, commonly used for graphical model structures, is used to find the Bayesian estimate of the edge matrix.  The edge matrix estimate is found by comparing the mean of the samples of edge matrices and determining if each off-diagonal element of the mean is greater than 0.5. If it is greater than 0.5, it is coded as an edge.  If the off-diagonal element of the mean is not greater than 0.5, it is coded as no edge. Models that are estimated to have no edges resulted in NaNs as MCC values.  The results are presented in Figures 1--3.  

The Spike Slab method has generally high specificity, compared to the models selected by the EBIC, StARS, and RIC methods.  The Spike Slab suffers in sensitivity for the 10\%, 5\%, and 2\% models, but the models selected by the EBIC, StARS, and RIC methods also suffer in sensitivity.  It is interesting to note that the EBIC selection method has been shown to perform well with the graphical lasso \citep{foygel_extended_2010}, but appears to suffer in performance when the graphical lasso is combined with the nonparanormal estimation method.  In particular, for the AR(1) and 10\% models for dimension $p=25$, the EBIC-selected model results in no edges.  When comparing the Spike Slab to the Bayesian Copula method, the Spike Slab has varying levels of success.  For $p=25$, the Bayesian Copula outperforms the Spike Slab, but for $p=50$ and $p=100$, the Spike Slab generally outperforms the Bayesian Copula method for the AR(1) and circle models.  Lastly, the Bayesian Copula method outperforms the Spike Slab for the sparsity percent models.  Overall, based on the MCC values, the Spike Slab performs similar to or better than the models selected by the EBIC, StARS, and RIC methods and similar to or better than Bayesian Copula for dimensions $p=50$ and $p=100$.

\begin{figure}[!htbp]
    \centering
    \includegraphics[width=.9\linewidth]{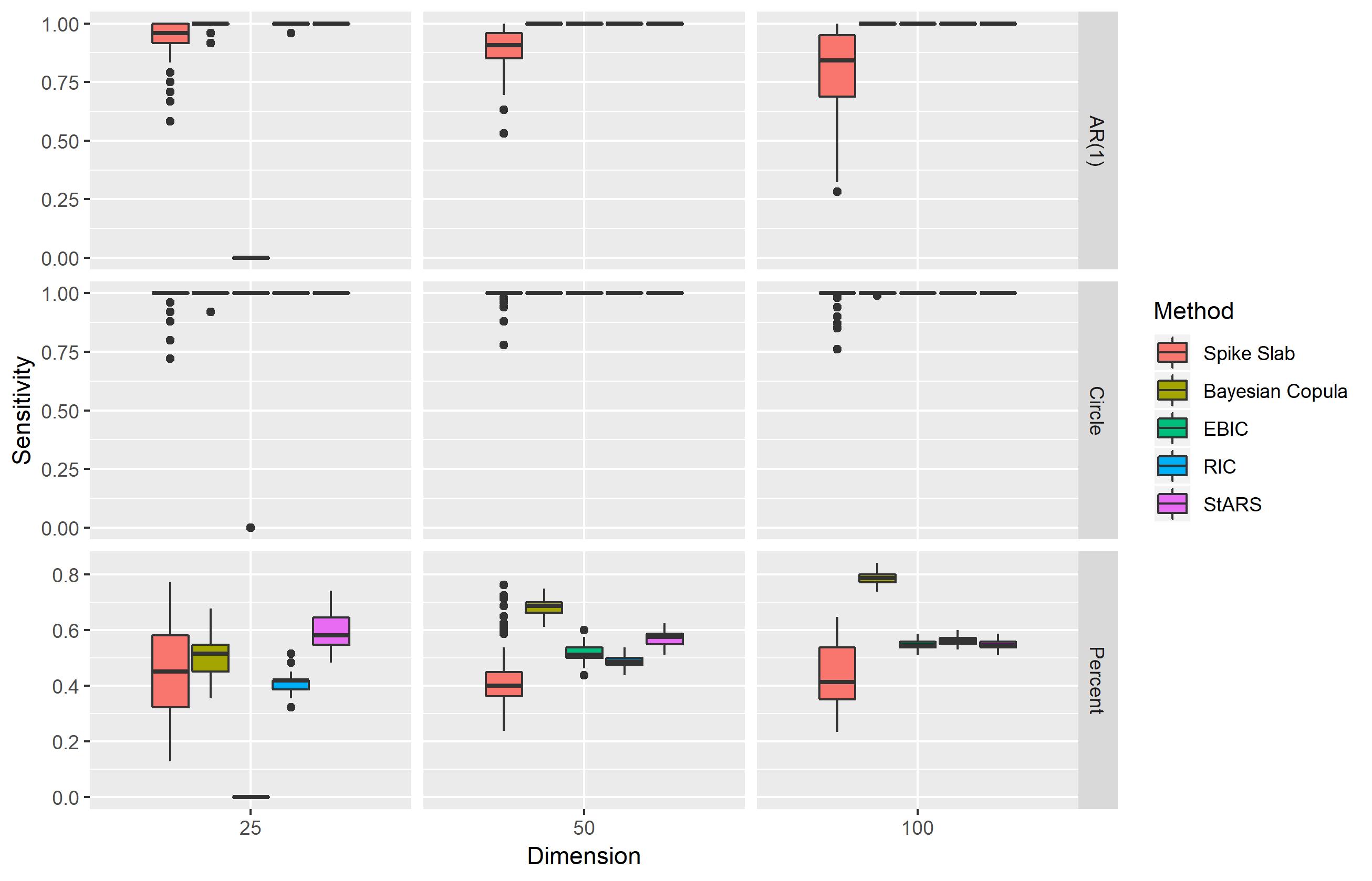}
    \caption{Boxplots of the sensitivity results for each of the methods for different structures of precision matrices. Percent refers to the 10\% model for dimension $p=25$, 5\% model for dimension $p=50$ and 2\% model for dimension $p=100$.}
   \vspace{1ex}
  \end{figure}

\begin{figure}[!htbp]
    \centering
    \includegraphics[width=.9\linewidth]{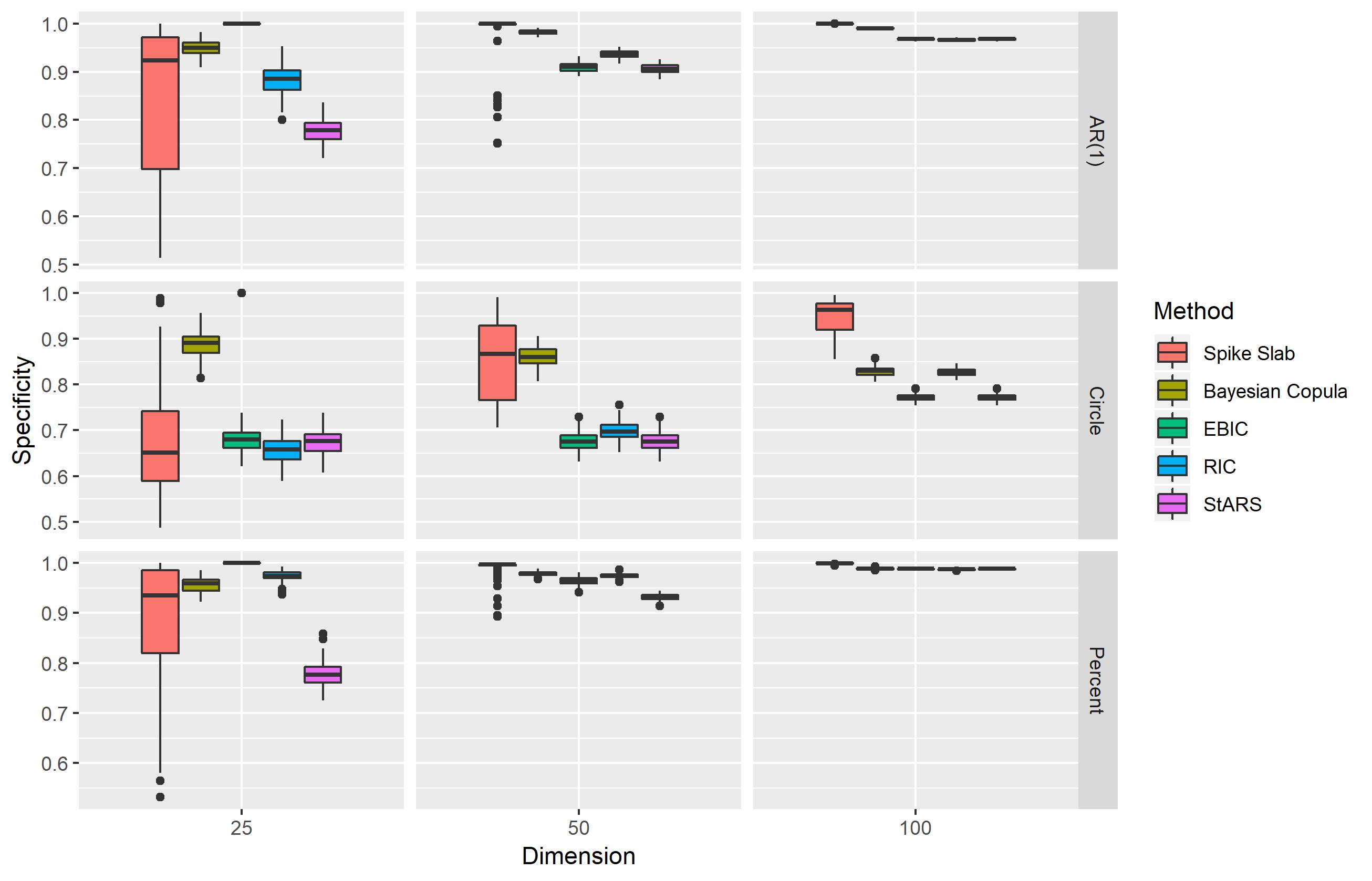}
    \caption{Boxplots of the specificity results for each of the methods for different structures of precision matrices. Percent refers to the 10\% model for dimension $p=25$, 5\% model for dimension $p=50$ and 2\% model for dimension $p=100$.}
   \vspace{1ex}
  \end{figure}
  
  \begin{figure}[!htbp]
    \centering
    \includegraphics[width=.9\linewidth]{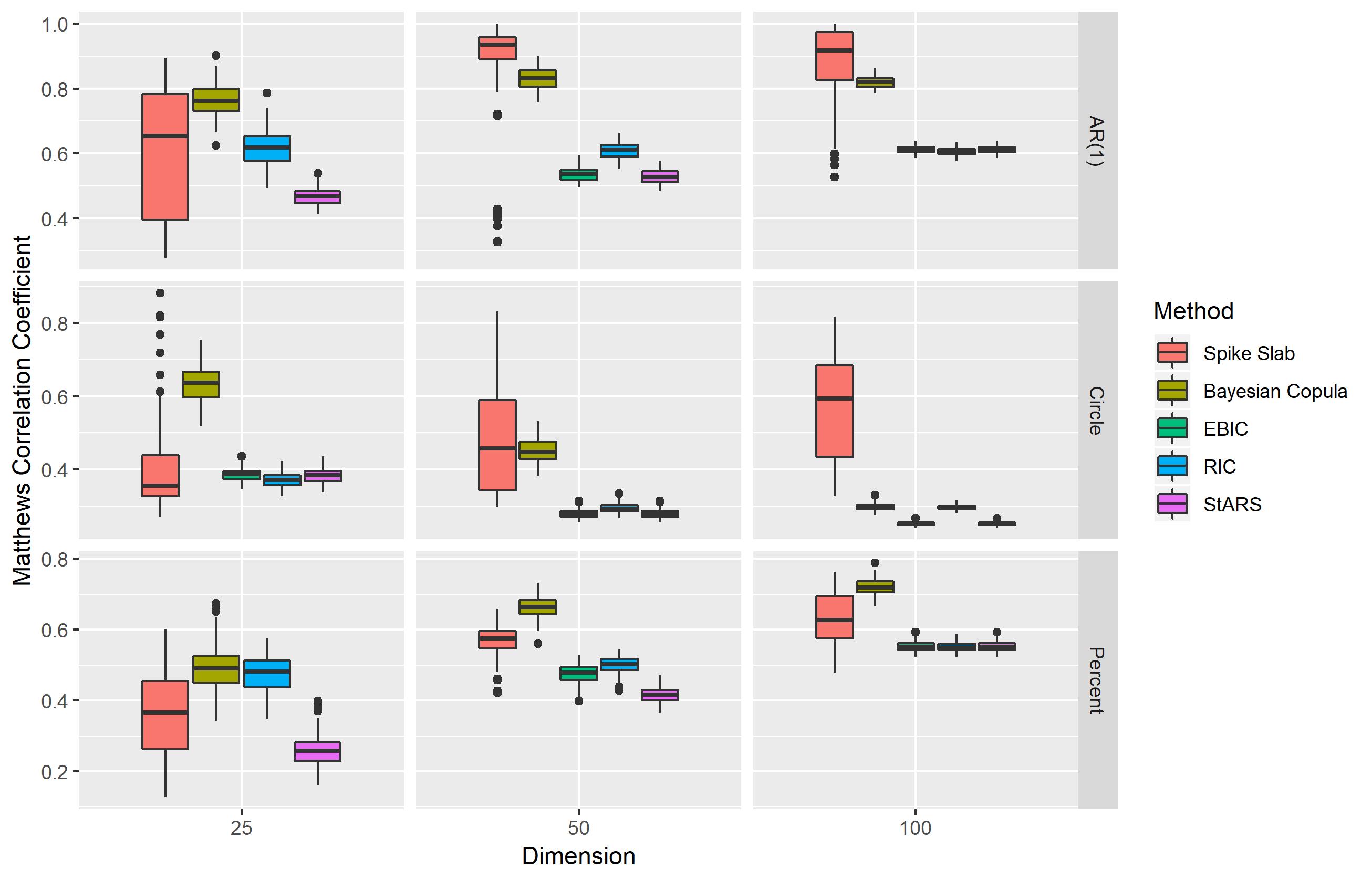}
    \caption{Boxplots of the Matthews correlation coefficient results for each of the methods for different structures of precision matrices. Percent refers to the 10\% model for dimension $p=25$, 5\% model for dimension $p=50$ and 2\% model for dimension $p=100$.}
   \vspace{1ex}
  \end{figure}

\section{Real Data Application}
We consider the data set based on the GeneChip (Affymetrix) microarrays for the plant \textit{Arabidopsis thaliana} originally referenced in  \citep{wille_sparse_2004}.  Since there are 118 microarrays, the sample size is $n=118$.  There are 39 genes from the isoprenoid pathway that are used.  For pre-processing, the expression levels for each gene, $x_i$ for $i = 1,\ldots, 118$, are log-transformed.  Additionally, for the Spike Slab method, the expression levels for each gene are converted to values between 0 and 1 using the equation $({x_i - \min(x_i)})/({\max(x_i)-\min(x_i)})$.  We study the associations among the genes using the Spike Slab method, the nonparanormal method of \cite{liu_nonparanormal:_2009}, as well as the Bayesian Copula method  \cite{mohammadi_bdgraph:_2017}.  These data are treated as multivariate Gaussian in the original analyses \citep{wille_sparse_2004}.  For the Spike Slab method, the final hyperparameter setting is chosen using the BIC method described in Section \ref{ChoiceParameters} and for the nonparanormal method of \cite{liu_nonparanormal:_2009}, 50 regularization parameters are used on an evenly spaced grid in the interval [0.16,1.2]. The three selection methods, RIC, EBIC, and StARS, are used with the default parameters in the {\tt huge} package. The nonparanormal model selected by EBIC result in no edges, so this model is not included in the comparison.  Bayesian model averaging is used for the posterior graph selection of the Bayesian Copula method using the {\tt BDgraph} package, in which it selects the graph with links for which the
estimated posterior probabilities are greater than 0.5.  The Spike Slab method can converge in about 29 minutes on a laptop for a given hyperparameter setting for these data.  The graphs are displayed in Figures 1--2. Plots are made with the {\tt circularGraph} function in {\tt MATLAB}. 

Our study shows that each of the methods leads to graphs with different levels of sparsity.  In particular, the Spike Slab and the nonparanormal using the RIC method lead to more sparsity  than the nonparanormal using the StARS method and the Bayesian Copula method.  The Spike Slab method results in 93 edges, the nonparanormal method using RIC results in 133 edges, and the nonparanormal method using StARS results in 209 edges.  The Bayesian Copula method resulted in 231 edges.  The Spike Slab model and the nonparanormal model selected with the RIC both capture some of the same edges, so these edges could be considered for further analysis.  Sparse models may aid in scientific exploration and interpretation. 

\begin{figure}[!htbp]
  \begin{subfigure}[b]{1\linewidth}
    \centering
    \includegraphics[width=.75\linewidth]{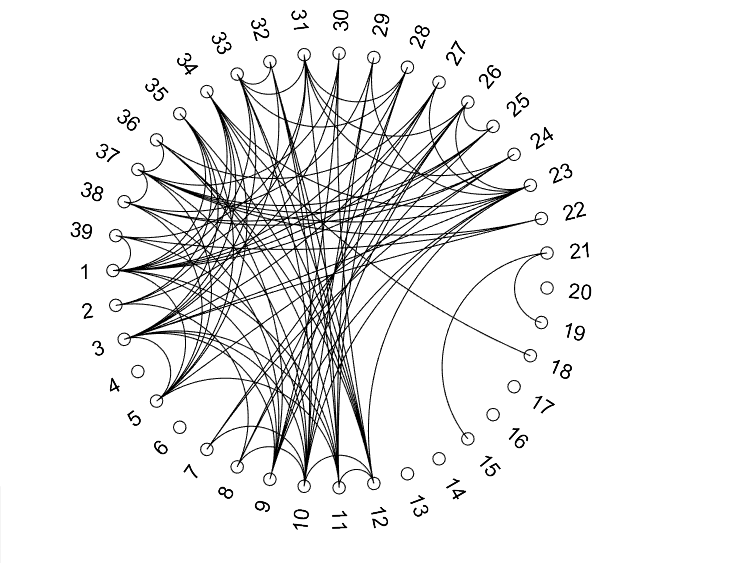}
    \caption{Spike Slab.}
   \vspace{3ex}
  \end{subfigure}
  \begin{subfigure}[b]{1\linewidth}
    \centering
    \includegraphics[width=.75\linewidth]{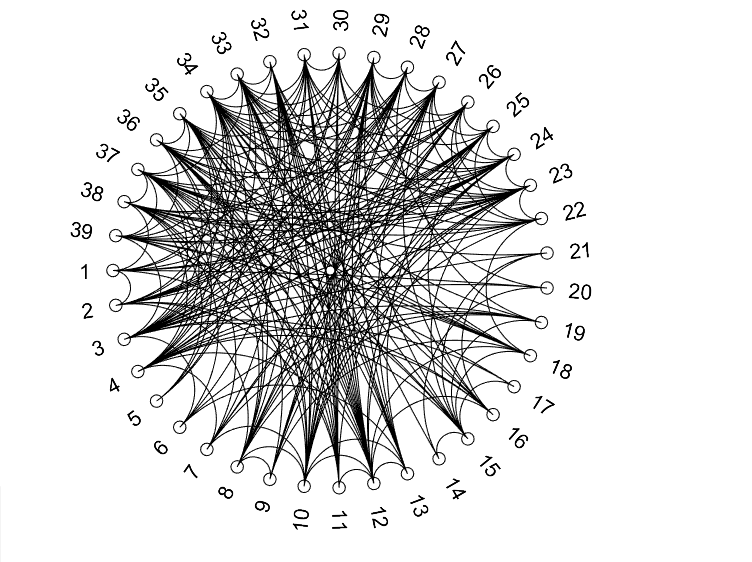}
    \caption{Bayesian Copula.}
    \vspace{3ex}
  \end{subfigure}
    \caption{Circular graphs illustrating the differences in edges between the methods using the microarray data set.}
\end{figure}

\begin{figure}[!htbp]
  \begin{subfigure}[b]{1\linewidth}
    \centering
    \includegraphics[width=.75\linewidth]{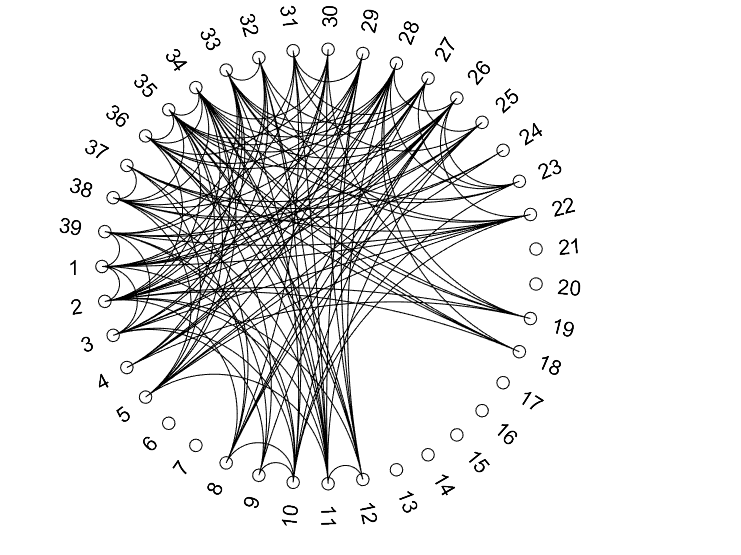}
    \caption{RIC.}
    \vspace{3ex}
  \end{subfigure}
    \begin{subfigure}[b]{1\linewidth}
    \centering
    \includegraphics[width=.75\linewidth]{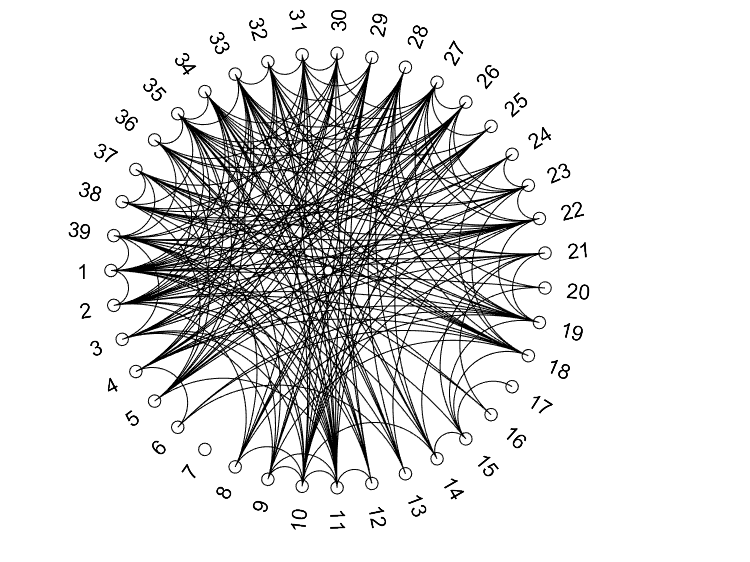}
    \caption{StARS.}
    \vspace{3ex}
  \end{subfigure}
  
  \caption{Circular graphs illustrating the differences in edges between the methods using the microarray data set.}
\end{figure}

\section{Discussion}
We have introduced a Bayesian method to construct graphical models for continuous data that do not rely on a normality assumption.  The method assumes the nonparanormal structure, that under some unknown monotone transformations on each component, the original observation vector reduces to a multivariate normal vector. The precision matrix of the transformed observations thus also determines the graphical structure of conditional independence of the original observations. We have considered a prior distribution on the underlying transformations through a finite random series of B-splines with increasing coefficients that are given a multivariate truncated normal prior.  The precision matrix of the transformed observations is given a spike-and-slab prior distribution. The procedure requires carefully considering identifiability restrictions. We have shown that certain linear constraints on the coefficients can give rise to identifiability. The advantage of using linear restrictions only is that the truncated multivariate normal structure on the vector of coefficients can be maintained under the identifiability restrictions. This allows us to use an efficient Gibbs sampler to compute the posterior distribution. We have shown that a suitably modified posterior distribution leads to posterior consistency of the mean and the variance of the transformed observations and the transformation functions using Euclidean distances on the mean and variance and the uniform pseudo-distance on a compact subset of the unit interval for the transformation functions. Since the transformations are learned through the marginal distributions of the component variables, their learning is largely independent of the learning of the precision matrix. Rather, the learning of the precision matrix is dependent on the transformations. The posterior consistency result we provide shows that, after a small tweaking of the posterior distribution of the transformations to comply with restrictions, for each component, the posterior distribution of the underlying transformation is consistent. In large samples, this ensures that the transformations are close to the unknown true transformations.  In addition, the posterior for the precision matrix is consistent, computed using the assumed consistent transformations.  

The Bayesian method, Spike Slab, appears to perform better than an earlier proposed empirical estimation method in the nonparanormal model at picking up edges that are significantly different from zero, thereby resulting in sparser models. Although it is not feasible to check for the nonparanormal distribution and therefore determine if our transformations improve on detecting the true transformation functions compared to the previous method, we believe that the use of the smooth and strictly increasing transformation functions that take into account for non-normality in combination with the prior on the precision matrix that incorporates sparsity improves on the goal of learning the structure of Gaussian graphical models when the data are continuous but not Gaussian.

\begin{supplement}
\sname{Supplement A}\label{suppA} 
\stitle{Proof of Consistency Theorems}
\sdescription{The supplement that includes the proof to the consistency theorems.}  
\slink[url]{https://github.com/jnj2102/BayesianNonparanormal}
\end{supplement}
\begin{supplement}
\sname{Supplement B}\label{suppB} 
\stitle{GitHub Repository: Bayesian Nonparanormal}
\slink[url]{https://github.com/jnj2102/BayesianNonparanormal}
\sdescription{The code used to run the methods described in this paper are available on GitHub.} 
\end{supplement}


\bibliographystyle{apalike}
\bibliography{Zotero.bib}

\begin{thebibliography}{}

\bibitem[Arbel et~al., 2013]{arbel_bayesian_2013}
Arbel, J., Gayraud, G., and Rousseau, J. (2013).
\newblock Bayesian optimal adaptive estimation using a sieve prior.
\newblock {\em Scandinavian Journal of Statistics}, 40(3):549--570.

\bibitem[Armagan et~al., 2013]{armagan_generalized_2013}
Armagan, A., B.~Dunson, D., and Lee, J. (2013).
\newblock Generalized double {Pareto} shrinkage.
\newblock {\em Statistica Sinica}, 23(1):119--143.

\bibitem[Baldi et~al., 2000]{baldi_assessing_2000}
Baldi, P., Brunak, S., Chauvin, Y., Andersen, C. A.~F., and Nielsen, H. (2000).
\newblock Assessing the accuracy of prediction algorithms for classification:
  an overview.
\newblock {\em Bioinformatics}, 16(5):412--424.

\bibitem[Banerjee et~al., 2008]{banerjee_model_2008}
Banerjee, O., El~Ghaoui, L., and d'Aspremont, A. (2008).
\newblock Model selection through sparse maximum likelihood estimation for
  multivariate {Gaussian} or binary data.
\newblock {\em Journal of Machine Learning Research}, 9:485--516.

\bibitem[Banerjee and Ghosal, 2014]{banerjee_posterior_2014}
Banerjee, S. and Ghosal, S. (2014).
\newblock Posterior convergence rates for estimating large precision matrices
  using graphical models.
\newblock {\em Electronic Journal of Statistics}, 8(2):2111--2137.

\bibitem[Banerjee and Ghosal, 2015]{banerjee_bayesian_2015}
Banerjee, S. and Ghosal, S. (2015).
\newblock Bayesian structure learning in graphical models.
\newblock {\em Journal of Multivariate Analysis}, 136:147 -- 162.

\bibitem[Berger and Barbieri, 2004]{berger_optimal_2004}
Berger, J.~O. and Barbieri, M.~M. (2004).
\newblock Optimal predictive model selection.
\newblock {\em The Annals of Statistics}, 32(3):870--897.

\bibitem[Bhattacharya et~al., 2015]{bhattacharya_dirichlet-laplace_2015}
Bhattacharya, A., Pati, D., Pillai, N.~S., and Dunson, D.~B. (2015).
\newblock Dirichlet-{Laplace} priors for optimal shrinkage.
\newblock {\em Journal of the American Statistical Association},
  110(512):1479--1490.

\bibitem[Carter et~al., 2011]{carter_constructing_2011}
Carter, C.~K., Wong, F., and Kohn, R. (2011).
\newblock Constructing priors based on model size for nondecomposable
  {Gaussian} graphical models: a simulation based approach.
\newblock {\em Journal of Multivariate Analysis}, 102(5):871--883.

\bibitem[Carvalho et~al., 2010]{carvalho_horseshoe_2010}
Carvalho, C.~M., Polson, N.~G., and Scott, J.~G. (2010).
\newblock The horseshoe estimator for sparse signals.
\newblock {\em Biometrika}, 97(2):465--480.

\bibitem[Choudhuri et~al., 2007]{choudhuri_nonparametric_2007}
Choudhuri, N., Ghosal, S., and Roy, A. (2007).
\newblock Nonparametric binary regression using a {Gaussian} process prior.
\newblock {\em Statistical Methodology}, 4(2):227--243.

\bibitem[Dahl et~al., 2005]{dahl_maximum_2005}
Dahl, J., Roychowdhury, V., and Vandenberghe, L. (2005).
\newblock Maximum likelihood estimation of {Gaussian} graphical models:
  numerical implementation and topology selection. {Technical} report.
\newblock University of California, Los Angeles.

\bibitem[Dahl et~al., 2008]{dahl_covariance_2008}
Dahl, J., Vandenberghe, L., and Roychowdhury, V. (2008).
\newblock Covariance selection for nonchordal graphs via chordal embedding.
\newblock {\em Optimization Methods and Software}, 23(4):501--520.

\bibitem[d'Aspremont et~al., 2008]{daspremont_first-order_2008}
d'Aspremont, A., Banerjee, O., and El~Ghaoui, L. (2008).
\newblock First-order methods for sparse covariance selection.
\newblock {\em SIAM Journal on Matrix Analysis and Applications}, 30(1):56--66.

\bibitem[de~Jonge and van Zanten, 2012]{de_jonge_adaptive_2012}
de~Jonge, R. and van Zanten, J. (2012).
\newblock Adaptive estimation of multivariate functions using conditionally
  {Gaussian} tensor-product spline priors.
\newblock {\em Electronic Journal of Statistics}, 6(0):1984--2001.

\bibitem[Dobra and Lenkoski, 2011]{dobra_copula_2011}
Dobra, A. and Lenkoski, A. (2011).
\newblock Copula {Gaussian} graphical models and their application to modeling
  functional disability data.
\newblock {\em The Annals of Applied Statistics}, 5(2A):969--993.

\bibitem[Foygel and Drton, 2010]{foygel_extended_2010}
Foygel, R. and Drton, M. (2010).
\newblock Extended {Bayesian} information criteria for {Gaussian} graphical
  models.
\newblock In {\em Advances in {Neural} {Information} {Processing} {Systems}
  23}, pages 604--612.

\bibitem[Friedman et~al., 2008]{friedman_sparse_2008}
Friedman, J., Hastie, T., and Tibshirani, R. (2008).
\newblock Sparse inverse covariance estimation with the graphical lasso.
\newblock {\em Biostatistics}, 9(3):432--441.

\bibitem[Ghosal and van~der Vaart, 2017]{ghosal_fundamentals_2017}
Ghosal, S. and van~der Vaart, A. (2017).
\newblock {\em Fundamentals of {Nonparametric} {Bayesian} {Inference}}.
\newblock Cambridge {Series} in {Statistical} and {Probabilistic} {Mathematics}
  (44). Cambridge University Press, Cambridge.

\bibitem[Giudici, 1999]{giudici_decomposable_1999}
Giudici, P. (1999).
\newblock Decomposable graphical {Gaussian} model determination.
\newblock {\em Biometrika}, 86(4):785--801.

\bibitem[Hájek et~al., 1999]{hajek_theory_1999}
Hájek, J., Šidák, Z., and Sen, P.~K. (1999).
\newblock {\em Theory of {Rank} {Tests}}.
\newblock Probability and {Mathematical} {Statistics}. Academic Press, Inc.,
  San Diego, CA, second edition.

\bibitem[Lenk and Choi, 2017]{lenk_bayesian_2017}
Lenk, P.~J. and Choi, T. (2017).
\newblock Bayesian {Analysis} of {Shape}-{Restricted} {Functions} using
  {Gaussian} {Process} {Priors}.
\newblock {\em Statistica Sinica}.

\bibitem[Letac and Massam, 2007]{letac_wishart_2007}
Letac, G. and Massam, H. (2007).
\newblock Wishart distributions for decomposable graphs.
\newblock {\em The Annals of Statistics}, 35(3):1278--1323.

\bibitem[Liu et~al., 2012]{liu_high-dimensional_2012}
Liu, H., Han, F., Yuan, M., Lafferty, J., and Wasserman, L. (2012).
\newblock High-dimensional semiparametric {Gaussian} copula graphical models.
\newblock {\em The Annals of Statistics}, 40(4):2293--2326.

\bibitem[Liu et~al., 2009]{liu_nonparanormal:_2009}
Liu, H., Lafferty, J.~D., and Wasserman, L.~A. (2009).
\newblock The nonparanormal: semiparametric estimation of high dimensional
  undirected graphs.
\newblock {\em Journal of Machine Learning Research}, 10:2295--2328.

\bibitem[Liu et~al., 2010]{liu_stability_2010}
Liu, H., Roeder, K., and Wasserman, L. (2010).
\newblock Stability approach to regularization selection ({StARS}) for high
  dimensional graphical models.
\newblock In {\em Advances in {Neural} {Information} {Processing} {Systems}
  23}, pages 1432--1440, USA.

\bibitem[Lu, 2009]{lu_smooth_2009}
Lu, Z. (2009).
\newblock Smooth optimization approach for sparse covariance selection.
\newblock {\em SIAM Journal on Optimization}, 19(4):1807--1827.

\bibitem[Lysen, 2009]{lysen_permuted_2009}
Lysen, S. (2009).
\newblock {\em Permuted inclusion criterion: a variable selection technique}.
\newblock PhD thesis, Publicly Accessible Penn Dissertations, 28.

\bibitem[Mazumder and Hastie, 2012]{mazumder_graphical_2012}
Mazumder, R. and Hastie, T. (2012).
\newblock The graphical lasso: new insights and alternatives.
\newblock {\em Electronic Journal of Statistics}, 6(0):2125--2149.

\bibitem[Meinshausen and Buhlmann, 2006]{meinshausen_high-dimensional_2006}
Meinshausen, N. and Buhlmann, P. (2006).
\newblock High-dimensional graphs and variable selection with the lasso.
\newblock {\em The Annals of Statistics}, 34(3):1436--1462.

\bibitem[Mohammadi et~al., 2017]{mohammadi_bayesian_2017}
Mohammadi, A., Abegaz, F., van~den Heuvel, E., and Wit, E.~C. (2017).
\newblock Bayesian modelling of {Dupuytren} disease by using {Gaussian} copula
  graphical models.
\newblock {\em Journal of the Royal Statistical Society: Series C (Applied
  Statistics)}, 66(3):629--645.

\bibitem[Mohammadi and Wit, 2015]{mohammadi_bayesian_2015}
Mohammadi, A. and Wit, E.~C. (2015).
\newblock Bayesian structure learning in sparse {Gaussian} graphical models.
\newblock {\em Bayesian Analysis}, 10(1):109--138.

\bibitem[Mohammadi and Wit, 2017]{mohammadi_bdgraph:_2017}
Mohammadi, R. and Wit, E.~C. (2017).
\newblock {BDgraph}: an {R} package for {Bayesian} structure learning in
  graphical models. {arXiv} preprint {arXiv}:1501.05108.

\bibitem[Mohammadi and Wit, 2019]{mohammadi_bdgraph:_2019}
Mohammadi, R. and Wit, E.~C. (2019).
\newblock {BDgraph}: {Bayesian} structure learning in graphical models using
  birth-death {MCMC}. {R} package version 2.57.

\bibitem[Pakman and Paninski, 2014]{pakman_exact_2014}
Pakman, A. and Paninski, L. (2014).
\newblock Exact {Hamiltonian} {Monte} {Carlo} for truncated multivariate
  {Gaussians}.
\newblock {\em Journal of Computational and Graphical Statistics},
  23(2):518--542.

\bibitem[Pitt et~al., 2006]{pitt_efficient_2006}
Pitt, M., Chan, D., and Kohn, R. (2006).
\newblock Efficient {Bayesian} inference for {Gaussian} copula regression
  models.
\newblock {\em Biometrika}, 93(3):537--554.

\bibitem[Rasmussen and Williams, 2006]{rasmussen_gaussian_2006}
Rasmussen, C.~E. and Williams, C. K.~I. (2006).
\newblock {\em Gaussian {Processes} for {Machine} {Learning}}.
\newblock Adaptive {Computation} and {Machine} {Learning}. MIT Press,
  Cambridge, Mass.

\bibitem[Rivoirard and Rousseau, 2012]{rivoirard_posterior_2012}
Rivoirard, V. and Rousseau, J. (2012).
\newblock Posterior concentration rates for infinite dimensional exponential
  families.
\newblock {\em Bayesian Analysis}, 7(2):311--334.

\bibitem[Rothman et~al., 2008]{rothman_sparse_2008}
Rothman, A.~J., Bickel, P.~J., Levina, E., and Zhu, J. (2008).
\newblock Sparse permutation invariant covariance estimation.
\newblock {\em Electronic Journal of Statistics}, 2(0):494--515.

\bibitem[Royston, 1982]{royston_algorithm_1982}
Royston, J.~P. (1982).
\newblock Algorithm {AS} 177: expected normal order statistics (exact and
  approximate).
\newblock {\em Applied Statistics}, 31(2):161.

\bibitem[Scheinberg et~al., 2010]{scheinberg_sparse_2010}
Scheinberg, K., Ma, S., and Goldfarb, D. (2010).
\newblock Sparse inverse covariance selection via alternating linearization
  methods.
\newblock In {\em Proceedings of the 23rd {International} {Conference} on
  {Neural} {Information} {Processing} {Systems} - {Volume} 2}, {NIPS}'10, pages
  2101--2109, USA. Curran Associates Inc.

\bibitem[Scheipl et~al., 2012]{scheipl_spike-and-slab_2012}
Scheipl, F., Fahrmeir, L., and Kneib, T. (2012).
\newblock Spike-and-slab priors for function selection in structured additive
  regression models.
\newblock {\em Journal of the American Statistical Association},
  107(500):1518--1532.

\bibitem[Shen and Ghosal, 2015]{shen_adaptive_2015}
Shen, W. and Ghosal, S. (2015).
\newblock Adaptive {Bayesian} procedures using random series priors: adaptive
  {Bayesian} procedures.
\newblock {\em Scandinavian Journal of Statistics}, 42(4):1194--1213.

\bibitem[Talluri et~al., 2014]{talluri_bayesian_2014}
Talluri, R., Baladandayuthapani, V., and Mallick, B.~K. (2014).
\newblock Bayesian sparse graphical models and their mixtures: sparse graphical
  modelling.
\newblock {\em Stat}, 3(1):109--125.

\bibitem[Uhler et~al., 2018]{uhler_exact_2018}
Uhler, C., Lenkoski, A., and Richards, D. (2018).
\newblock Exact formulas for the normalizing constants of {Wishart}
  distributions for graphical models.
\newblock {\em The Annals of Statistics}, 46(1):90--118.

\bibitem[van~der Vaart and van Zanten, 2007]{van_der_vaart_bayesian_2007}
van~der Vaart, A. and van Zanten, H. (2007).
\newblock Bayesian inference with rescaled {Gaussian} process priors.
\newblock {\em Electronic Journal of Statistics}, 1(0):433--448.

\bibitem[Wang, 2012]{wang_bayesian_2012}
Wang, H. (2012).
\newblock Bayesian graphical lasso models and efficient posterior computation.
\newblock {\em Bayesian Analysis}, 7(4):867--886.

\bibitem[Wang, 2015]{wang_scaling_2015}
Wang, H. (2015).
\newblock Scaling it up: stochastic search structure learning in graphical
  models.
\newblock {\em Bayesian Analysis}, 10(2):351--377.

\bibitem[Wang and Li, 2012]{wang_efficient_2012}
Wang, H. and Li, S.~Z. (2012).
\newblock Efficient {Gaussian} graphical model determination under
  {G}-{Wishart} prior distributions.
\newblock {\em Electronic Journal of Statistics}, 6(0):168--198.

\bibitem[Wille et~al., 2004]{wille_sparse_2004}
Wille, A., Zimmermann, P., Vranová, E., Fürholz, A., Laule, O., Bleuler, S.,
  Hennig, L., Prelić, A., von Rohr, P., Thiele, L., Zitzler, E., Gruissem, W.,
  and Bühlmann, P. (2004).
\newblock Sparse graphical {Gaussian} modeling of the isoprenoid gene network
  in \textit{{Arabidopsis} thaliana}.
\newblock {\em Genome Biology}, 5(11):R92--R92.

\bibitem[Witten et~al., 2011]{witten_new_2011}
Witten, D.~M., Friedman, J.~H., and Simon, N. (2011).
\newblock New insights and faster computations for the graphical lasso.
\newblock {\em Journal of Computational and Graphical Statistics},
  20(4):892--900.

\bibitem[Wong et~al., 2003]{wong_efficient_2003}
Wong, F., Carter, C.~K., and Kohn, R. (2003).
\newblock Efficient estimation of covariance selection models.
\newblock {\em Biometrika}, 90(4):809--830.

\bibitem[Yuan and Lin, 2007]{yuan_model_2007}
Yuan, M. and Lin, Y. (2007).
\newblock Model selection and estimation in the {Gaussian} graphical model.
\newblock {\em Biometrika}, 94(1):19--35.

\bibitem[Zhao et~al., 2015]{zhao_huge:_2015}
Zhao, T., Li, X., Liu, H., Roeder, K., Lafferty, J., and Wasserman, L. (2015).
\newblock huge: high-dimensional undirected graph estimation.
\newblock R package version 1.2.7.

\end{thebibliography}

\section*{Acknowledgement}
Research of the first author is supported by the National Science Foundation (NSF) Graduate Research Fellowship Program Grant No. DGE-1252376, the National Institutes of Health (NIH) training grant GM081057 and NSF grant DMS-1732842. Research of the second author is partially supported by NSF grant DMS-1510238. 

\end{document}